\pdfoutput=1
 \documentclass[a4paper,fleqn,usenatbib,usedcolumn]{mnras}

 \usepackage[british]{babel}             % British English hyphenation
 \usepackage{newtxtext}                  % Good fonts
 \usepackage[slantedGreek]{newtxmath}    %   "    "   (slanted Greek)
 \let\la=\lesssim % for less than similar from newtxmath, not \la from mnras.cls
\usepackage[T1]{fontenc}
\usepackage{ae,aecompl}

\usepackage{graphicx}	% Including figure files
\usepackage{amsmath}	% Advanced maths commands
\usepackage{amssymb}	% Extra maths symbols

\RequirePackage{fix-cm} % suppress some of the font warnings

\newcommand{\flame}{\texttt{flame}}

\title[Data Reduction for Near-Infrared and Optical Spectroscopy]{\emph{Flame}: A Flexible Data Reduction Pipeline \\
for Near-Infrared and Optical Spectroscopy}

\author[Belli, Contursi, \& Davies]{
Sirio Belli, Alessandra Contursi, and Richard I. Davies
\\
Max-Planck-Institut f\"ur Extraterrestrische Physik (MPE), Giessenbachstr. 1, D-85748 Garching, Germany\\
}

\pubyear{2017}

\begin{document}
\label{firstpage}
\pagerange{\pageref{firstpage}--\pageref{lastpage}}
\maketitle

\begin{abstract}
We present \flame, a pipeline for reducing spectroscopic observations obtained with multi-slit near-infrared and optical instruments. Because of its flexible design, \flame\ can be easily applied to data obtained with a wide variety of spectrographs. The flexibility is due to a modular architecture, which allows changes and customizations to the pipeline, and relegates the instrument-specific parts to a single module. At the core of the data reduction is the transformation from observed pixel coordinates $(x,y)$ to rectified coordinates $(\lambda,\gamma)$. This transformation consists in the polynomial functions $\lambda(x,y)$ and $\gamma(x,y)$ that are derived from arc or sky emission lines and slit edge tracing, respectively. The use of 2D transformations allows one to wavelength-calibrate and rectify the data using just one interpolation step. Furthermore, the $\gamma(x,y)$ transformation includes also the spatial misalignment between frames, which can be measured from a reference star observed simultaneously with the science targets. The misalignment can then be fully corrected during the rectification, without having to further resample the data. Sky subtraction can be performed via nodding and/or modeling of the sky spectrum; the combination of the two methods typically yields the best results. We illustrate the pipeline by showing examples of data reduction for a near-infrared instrument (LUCI at the Large Binocular Telescope) and an optical one (LRIS at the Keck telescope).
\end{abstract}


%%%%%%%%%%%%%%%%%%%%%%%%%%%%%%%%%%%%%%%%%%%%%%%%%%

%%%%%%%%%%%%%%%%% BODY OF PAPER %%%%%%%%%%%%%%%%%%

% -------------------------------------------------
%  					INTRODUCTION
% -------------------------------------------------

\section{Introduction}
\label{sec:introduction}

Given the increasing complexity of modern instrumentation, observational astronomers today rarely develop their own software for data reduction. Typically, automated pipelines written specifically for each instrument are used. These pipelines can be very sophisticated, and are often tailored to the properties and features of the instrument. While this approach can lead to data reduction of high quality, it also presents significant limitations. First, such pipelines often lack flexibility and are difficult to adapt to specific scientific needs that differ from the intended use. Second, users typically treat them as ``black boxes'', without a proper understanding of the details involved in the data reduction, which leads to a poor understanding of the final data. Third, they cannot be used to reduce data obtained with one of the many existing instruments that lack an official pipeline.

These issues are particularly important for near-infrared spectroscopy, which has seen a rapid growth in recent years due to the development of sensitive detectors and complex multi-object instruments. Near-infrared observations are particularly demanding in terms of data reduction quality because of strong, variable emission from the sky, and significant systematics in the detectors. Observational techniques such as nodding and dithering are effective at mitigating these problems, but introduce further data reduction issues because of the extra resampling step that is required for proper alignment.

While near-infrared and optical observations differ in some of the details, they require the same basic data reduction steps. Furthermore, there is not a clear transition between the two regimes: near-infrared detectors replace CCDs for wavelengths beyond $\sim 1\mu m$, but OH emission lines, which dictate some of the main differences in the data reduction techniques, start dominating the sky spectrum at $\sim 7000\AA$, and decrease in strength and number beyond $\sim 2\mu m$. In practice, the details of each observation will determine which data reduction techniques to apply.

For these reasons we have developed \flame, a flexible data reduction pipeline that can be easily applied to any infrared or optical data taken with multi-slit instruments. In designing \flame\ we tried to achieve the simplest implementation that would follow these guiding principles:
\begin{itemize}
  \item Have a highly modular design that can be easily modified.
  \item With the exception of the initialization, where the options and parameters are set, the pipeline must be instrument-independent.
  \item Take full advantage of the reference star that is commonly placed on one of the slits in order to monitor pointing, seeing, and transmission.
  \item Deliver science-quality results, particularly for sky-limited observations of faint targets.
  \item Resample the data only once throughout the reduction in order to limit the impact of correlated noise.
  \item Provide a partial reduction on short timescales, so that data can be checked while observing.
\end{itemize}

We illustrate the pipeline by showing examples of data reduction for two of the instruments that are currently supported: LRIS, an optical spectrograph at the Keck telescope, and LUCI, a near-infrared spectrograph at the Large Binocular Telescope (LBT). Extending the range of supported instruments is relatively easy: it is sufficient to write a new initialization module that sets the required parameters and settings, and transforms the data into the required format, if needed.

The pipeline is written in IDL and is distributed\footnote{\url{https://github.com/siriobelli/flame}} with a GPL license. Here we give an overview of its design, focusing on the issues of data reduction that are more relevant for scientific observations, in particular for faint sources. A user manual explaining the technical details of the pipeline is also available.

We start by describing the main characteristics of multi-object spectroscopic observations in Section \ref{sec:observations}, where we also discuss the calibration data and the nodding and dithering methods. We then explain in detail the steps followed by \flame\ to reduce a set of data in Section \ref{sec:datareduction}. In Section \ref{sec:performance} we discuss two examples of data reduction: one in the near-infrared and one in the optical. Finally, we summarize the scope and methods of the pipeline in Section \ref{sec:summary}.

% -------------------------------------------------
%  					SPECTROSCOPIC OBSERVATIONS
% -------------------------------------------------

\section{Spectroscopic Observations}
\label{sec:observations}

In this section we briefly describe the typical steps involved in spectroscopic observations with multi-object instruments, noting in particular how they affect the data reduction methods.

\subsection{Science Observations}

\begin{figure}
  \centering
	\includegraphics[width=\columnwidth]{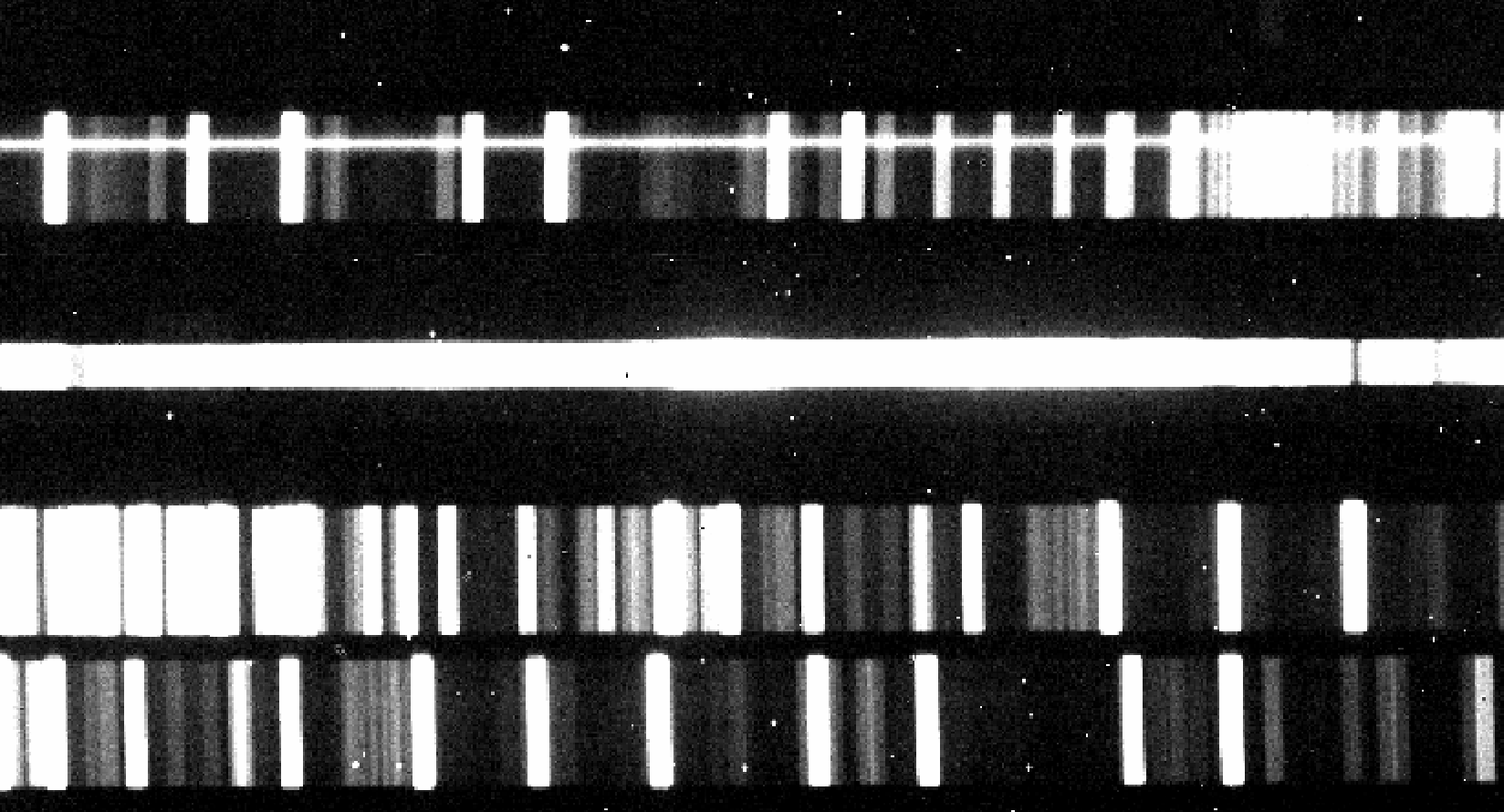}
   \caption{A small section of a raw LUCI frame. Wavelength runs from left to right, while the spatial direction is vertical. In the top slit the trace from the reference star is visible, together with bright vertical emission lines from the sky. The horizontal stripe under the top slit is the trace left by a bright star through the alignment box. The bottom two slits contain faint science targets, not visible in individual exposures, and are dominated by the sky emission.}
   \label{fig:example_frame}
\end{figure}

We assume that the scientific observations are taken using multi-object slitmasks, with the wavelength direction running horizontally and the spatial direction vertically. Each slit produces a 2D spectrum with a different wavelength range, which is set by the characteristics of the instrument and the horizontal position of the slit on the mask. Figure \ref{fig:example_frame} shows an example of a raw frame from near-infrared observations with the LUCI spectrograph.

Most of the slits are centered on the scientific targets, but it is good practice to reserve at least one slit to a relatively bright star which can be used to track seeing, transmission, and pointing, especially if the targets are faint or extended. Such \emph{reference star} plays a fundamental role in \flame, as will be clear below. In most cases, together with the science slits there are also wider slits, called ``alignment boxes'', centered on very bright stars, that are used to align the slitmask before the observations. If needed, alignment boxes can also be used to track the vertical offset and transmission.

Longslit observations are treated by \flame\ as a special case of a slitmask with only one slit.

\subsection{Nodding and Dithering}
\label{sec:nodding}

In the near-infrared it is common practice to offset, or ``nod'', the telescope pointing between successive frames. This allows to remove large part of the systematic effects due to both the sky and the detector by simply subtracting frames taken at different pointings. The pipeline supports the common 2-point nodding scheme, also called $A-B$, where $A$ and $B$ represent two successive frames taken at different nod positions.

\begin{figure}
  \centering
	\includegraphics[width=\columnwidth]{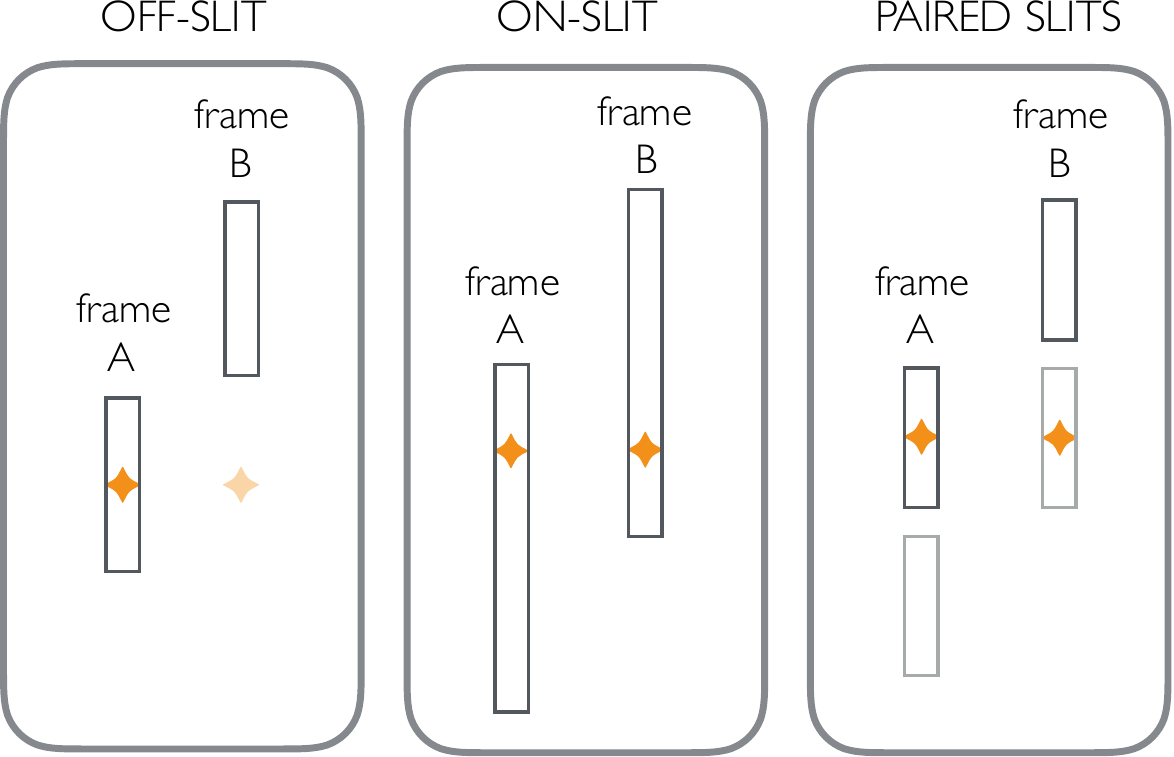}
   \caption{Types of 2-point nod patterns. \emph{Off-slit}: the target is observed in frame $A$, and the sky is observed in frame $B$. In this case, the target is not observed at all for half of the observing time. \emph{On-slit}: the target is observed in frame $A$, and is also observed in frame $B$ but at a different spatial position along the slit. \emph{Paired slits}: the target is observed in frame $A$, and the sky is observed in frame $B$; however the target in frame $B$ falls on a different slit (shown in lighter gray). For each of these nod patterns, small-scale dithering can be added by perturbing the pointing at each frame.}
   \label{fig:nodding_diagram}
\end{figure}

For multi-object spectroscopic observations we can identify three types of $A-B$ nodding, which are illustrated in Figure \ref{fig:nodding_diagram}. The simplest case is the \emph{off-slit} nod: the target is placed at the center of the slit in the $A$ frame, while is outside the slit in the $B$ frame, when blank sky is observed instead. However, this means that the exposure time available on the target is half of the total observing time. The \emph{on-slit} nod solves this problem: the target is always in the slit, but its spatial position in the $A$ frame is different from that in the $B$ frame. By analyzing both the $A-B$ and the $B-A$ frame differences, one can obtain a sky-subtracted spectrum of the target at both nod positions. The disadvantage of this pattern is that the slit needs to be longer to avoid subtracting the object from itself. Finally, another possible scheme is the \emph{paired slits} nod, where the object is placed on a different slit in the $B$ frame. This is particularly useful when each slit on the mask needs to have a different position angle, which makes the on-slit nod impossible.

Since in the near-infrared the sky OH lines are highly variable, the sky spectrum in the $A$ frames is never exactly identical to that in the $B$ frames. This can lead to strong residuals in the $A-B$ frame difference. With on-slit and paired slits nodding this effect is mitigated, because the final stack is a combination of $A-B$ and $B-A$ and the residuals are canceled (see Section \ref{sec:combine} for details on how \flame\ combines the frames for the various nod patterns). When using the off-slit nod, however, the target is never observed in the $B$ frames, which means that the residuals will strongly affect the final stack. In any case, additional processing of the data can help improve the sky subtraction. In \flame, a model of the sky is constructed in every frame using the pixels that are in empty regions of the slit. This is the standard method for optical observations and is discussed in detail in Section \ref{sec:skysub}. Other possibilities include to scale the different OH bands in the sky spectrum according to the algorithm by \citet{davies07}, or to use the principal component analysis technique developed by \citet{soto16}. These methods are not currently supported by \flame\ but can in principle be easily added.

Spatial dithering, which consists of varying the pointing by small amounts at each frame, is another common strategy employed in near-infrared observations. Dithering improves spatial sampling and mitigates the effect of bad pixels. In practice dithering is achieved by slightly perturbing the nod positions; a common pattern that includes both nodding and dithering is $ABA'B'$, where $A'$ and $B'$ are only slightly offset from $A$ and $B$, respectively.  In order to take full advantage of sub-pixel dithering, great care must be taken when resampling and aligning the frames.

Finally, the position of the target on the detector can also vary from frame to frame because of drifting due to flexure or imperfect guiding. This effect is noticeable in most instruments when the exposure time is sufficiently long. Drifting can be effectively considered as unintended dithering, and can be fully corrected by the data reduction, as long as it can be accurately measured.

While the nodding and dithering positions are in principle recorded in the FITS header of each frame, there is often a non-negligible difference between the offset commanded to the telescope and the one actually performed. Furthermore, the drifting of the target on the detector cannot be recovered from the information in the header. For these reasons having a reference star on one of the slits is critical for an accurate measurement of the true pointing at each frame, and for a correct handling of nodding, dithering, and drifting.

\subsection{Calibrations}

The types of calibrations needed to properly reduce the data depend on the specific observational setup used (wavelength range, exposure time, type of detector, etc). Here we discuss the calibrations that are implemented in \flame. Each calibration is optional; the pipeline does not \emph{require} any of them to function. The freedom - and responsibility - to choose which calibrations to use is entirely left to the user.

\subsubsection{Dark Frames}

Dark frames are taken without exposing the detector to any source of light, and using an integration time that matches that of the observations to be calibrated. The master dark frame is then subtracted from the scientific frames to account for the amount of flux that is measured by the detector but is not due to an astronomical source.

The flux measured in a dark frame can be decomposed into a part that scales linearly with time (the dark current), and a constant offset (the bias). This means that it is possible to combine a bias and a dark frame to obtain the calibrations appropriate for any arbitrary exposure time. \flame\ does not perform such combination, but the user can easily build the correct dark frames and feed them to the pipeline.

Dark subtraction is automatically accounted for when applying the $A-B$ subtraction, thus dark frames are not needed when nodding. However, they are still useful for the identification of hot pixels (see Section \ref{sec:calibrations}).

\subsubsection{Flat Fields}

Flat fields are obtained by illuminating evenly the detector, and are used to remove from the data unwanted features due to instrumental effects. There are, in practice, many different ways to obtain a flat field: the source of the emission can be artificial (i.e., an incandescent light bulb in the dome or inside the instrument) or natural (most often the sky at twilight); the light can be dispersed or not; the slitmask can be inserted in the focal plane or not; and so on.

From the point of view of the data reduction there is a better way to classify flat field calibrations, which is according to how they are used, rather than what observational setup was adopted in taking the data. In \flame\ the flat fields are divided into three types:
\begin{enumerate}
\item The \emph{pixel flat field} is used to correct for the pixel-to-pixel variation in sensitivity. For this purpose it is important to obtain a smooth illumination of the full detector. This is typically done using an incandescent lamp, and sometimes using a dispersion element to further smooth the illumination. The unavoidable large-scale variations in the illumination are removed via polynomial fitting, and the remaining flat field is normalized to unity. This represents a local correction to the gain value, which is different for each pixel, and can also depend on wavelength, particularly for near-infrared detectors. The pixel flat field is also used to detect dead pixels.
\item The \emph{illumination flat field} corrects for the non-uniform illumination of the slit along the spatial direction (while for the spectral direction standard star observations should be used, as discussed below). Ideally, the illumination flat field is obtained observing the sky at twilight with the same filter and dispersing element as the science observations; the data are then smoothed and the spectral variation removed. Using dome flats to derive an illumination correction is not always advisable because the optical path followed by the lamp light can be very different from that of the light from an astronomical source. An alternative way to derive the illumination correction is to use bright sky emission lines from the science observations, as discussed in Section \ref{sec:illumcorr}; in this case the illumination flat field is not needed.
\item Finally, the \emph{slit flat field} is used to identify the slits and trace their edges. While a lamp flat is often used for this purpose, the illumination does not need to be flat nor smooth, as long as the slit edges are recognizable. Good results can be obtained, for example, using arc lamps, or even using the science observations, if relatively bright continuum or line emission from the sky is present (see Section \ref{sec:slitid}).
\end{enumerate}

\subsubsection{Arcs}

Arcs calibrations are obtained with the same instrumental setup (filter, dispersing element, slitmask) as the observations, but with the source replaced by a lamp containing a gaseous mixture (e.g., Ne, Xe, Ar, etc.) whose spectrum consists of a number of bright emission lines of known wavelength. These lines can then be used to wavelength-calibrate the science observations.

Arcs are not always needed: if bright sky emission lines are present in the science observations, these can be used to obtain a wavelength solution. This method is in fact preferable because not affected by potential misalignments between arcs and science frames caused by flexure. Sky lines are generally sufficiently strong redwards of $7000\AA$, and data obtained in the red or near-infrared part of the spectrum will usually not require arcs calibrations.
However, even at these wavelengths there are cases where it may be necessary to use arcs: for example in the $K$ band, where bright OH lines are relatively sparse; or when the exposure times are short and the sky lines are not bright enough for accurate measurements.

\subsection{Flux Calibration and Telluric Correction}

The light emitted by an astronomical source goes through the atmosphere, the optics and the instrument components before hitting the detector. Each of these steps introduces a spectral signature, which needs to be removed from the final data.

The illumination flat field can, in principle, be used to remove the spectral response of the instrument. However, this works only when using a calibration lamp with a spectrum that is exactly flat. Furthermore, flat fielding cannot remove the telluric absorption due to the atmosphere, which is particularly strong in the near-infrared.

A better approach is to use the observation, at an airmass matching that of the science observations, of a standard star for which the intrinsic spectrum is known. This can be either a theoretical spectrum or a previously observed and calibrated spectrum. For near-infrared observations, \citet{vacca03} provide a method to obtain the intrinsic spectrum of any A0 star. The sensitivity can then be calibrated as a function of wavelength by dividing the expected flux by the observed flux of the star. This relative flux calibration will also automatically correct for the telluric absorption.

If the brightness of the standard star is known, then an absolute flux calibration can also be performed in the same step. Another possibility is to derive only a relative flux calibration from the standard star, and use the \emph{reference} star to measure the scaling needed to obtain an absolute calibration. This method usually yields a better calibration since slit loss, transmission conditions, and airmass are the same for the reference star and the science observations, but of course requires the use of a reference star of known brightness.

Flux calibration is not explicitely performed in \flame. This is, however, easy to achieve once the observations of the standard star are reduced in the same way as the science observations, and an intrinsic spectrum of the star is available.

% -------------------------------------------------
%  					DATA REDUCTION STEPS
% -------------------------------------------------

\section{Data Reduction Steps}
\label{sec:datareduction}

In order to ensure a high degree of modularity and flexibility, the design of \flame\ consists of a series of IDL routines, each handling a relatively simple task. The only argument taken by these routines is a structure containing all the information relevant for a data set, including the input, instrument properties, and settings. The data reduction is then entirely controlled via a sequence of IDL commands saved in a driver file, of which we show an example in Appendix \ref{sec:driverfile}.

We can conceptually divide the routines that make up \flame\ into four parts:
\begin{enumerate}
  \item \emph{Initialization}: the inputs are set, the only instrument-specific routine is run, and optional settings can be tweaked;
  \item \emph{Operations on frames}: diagnostics of the observing conditions are calculated, frames are calibrated, slits are identified and cutouts extracted;
  \item \emph{Calculation of the coordinate transformations}: using the slit edges, the position of the reference star, and the emission lines from either arcs or sky, the transformations between the observed pixel coordinates and the rectified spectral and spatial coordinates are calculated for each cutout;
  \item \emph{Operations on slits}: illumination correction is calculated and applied, sky emission is modeled and subtracted, cutouts are rectified and stacked, a data quality check is performed, and the 1D spectra are extracted.
\end{enumerate}
Here we describe in some detail the individual routines that compose each of these four parts. It should be clear from the section headings which IDL routine among the ones listed in Appendix \ref{sec:driverfile} is being discussed (e.g., ``Slit Identification'' refers to the routine \texttt{flame\_slitid}).

\subsection{Initialization}
\label{sec:initialize}

The initialization routine takes the inputs specified by the user and outputs a data structure containing all the information needed for the data reduction.

The only required input is the name of an ASCII file containing the list of science frames to be reduced. Further optional inputs include similar ASCII files for the calibration frames, the approximate pixel coordinate (or coordinates, if nodding) of the reference star trace, and settings for long-slit observations. It is also possible to flag a specific slit as the only one to be reduced, thus substantially decreasing the execution time, which can be particularly useful while observing.

After all the inputs are set, the initialization is run. One of the main tasks of this module is to translate the specific configuration of the instrument (typically obtained from the FITS header) into generic settings that can be understood by \flame. For this reason, it is necessary to use a different initialization module for each instrument. For example, the names of gratings and cameras and filters need to be translated into a spatial pixel scale and wavelength range. Other values that need to be set at this stage are the gain, readout noise, linearity correction parameters, approximate spectral resolution, and so on.

During the initialization the various settings that control the data reduction are tweaked to the instrumental configuration; however, the user has then complete control over these settings. For example, near-infrared observations most often relies on sky lines for wavelength calibration, and the LUCI initialization sets this method by default, but users are free to change this setting and supply arcs calibrations. A complete list of settings is reported in the \flame\ user manual; these include virtually all parameters that can be tweaked and that are involved in any of the data reduction steps, such as the polynomial degree used for the wavelength solution, or whether cosmic ray cleaning should be performed.

Some instruments output raw data in formats that are significantly different from that assumed by \flame. The most common case is when the wavelength direction runs along the vertical direction of the detector instead of the horizontal one. Or the raw data may consist of FITS files with multiple extensions, for example when more than one detectors are used. In all these cases, the initialization routine transforms the raw data into the standard format accepted by \flame.

Currently, \flame\ includes initialization modules for the LUCI spectrograph at LBT and the LRIS and MOSFIRE spectrographs at Keck. Users can easily add support for a new instrument by writing the corresponding initialization module. The user manual explains in detail the steps required to initialize the data reduction. We also provide a generic initialization module that can be used as a template.

\subsection{Operations on frames}

\subsubsection{Diagnostics and Quick Stack}
\label{sec:diagnostics}

\begin{figure}
  \centering
  \includegraphics[width=0.9\columnwidth]{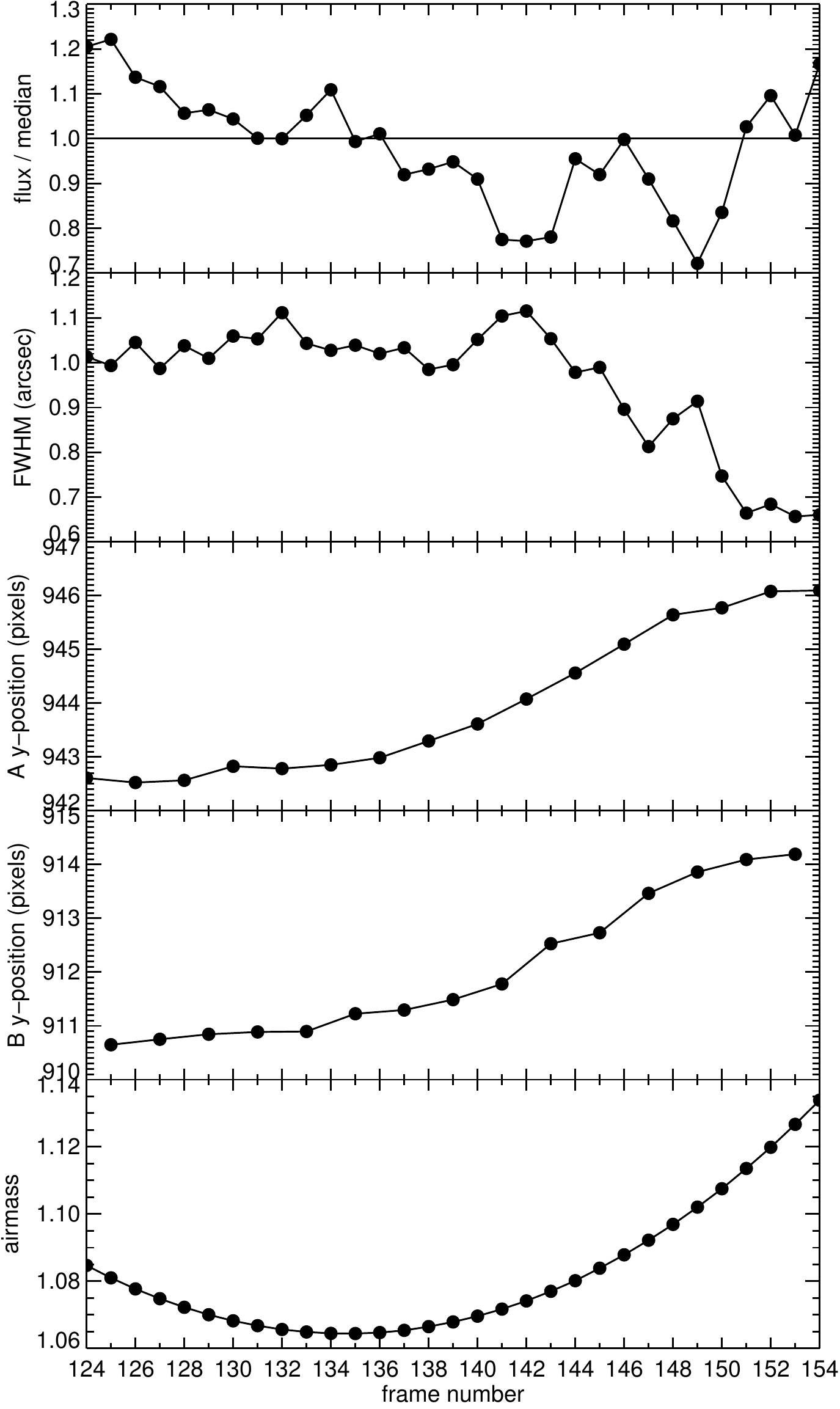}
  \caption{Example of diagnostic plots output by \flame\ for observations obtained with $A-B$ nodding. From top to bottom, the panels show: relative variation in the integrated flux of the reference star; spatial FWHM of the reference star trace; vertical position of the reference star trace for both the $A$ and $B$ positions; airmass. These observations were taken with LBT/LUCI in the $H$ band, for a total exposure time of 155 minutes. In the first half of the observations, airmass and seeing are stable, the pointing is only mildly varying, yet the flux shows a steady decline, perhaps due to slit losses caused by imperfect guiding. Then, when the airmass starts increasing, a large drift in the vertical position is visible, probably due to flexure effects. This is, however, counteracted by an improvement in the seeing, and consequently an increase in the flux.}
   \label{fig:diagnostics}
\end{figure}

If a reference star is present on the slitmask, its trace is used in this step to derive useful diagnostics of the observing conditions. Starting with the approximate position input by the user, the trace of the reference star is automatically identified in each frame, and its profile is extracted using a narrow range in $x$ to avoid blurring due to geometric distortions. The flux, full width at half maximum (FWHM), and center of the profile are then measured via Gaussian fitting. These quantities are saved and used to generate diagnostic plots such as the one shown in Figure \ref{fig:diagnostics}.

There are many important uses for such measurements of the observing conditions. First, this step can be run in real time while observing to provide critical information on the quality of the data. Second, when reducing the data it is often necessary to identify the frames with the poorest conditions and possibly exclude them from the final reduction. Finally, some of the measurements derived in this phase (most importantly, the vertical position) will be used in later steps of the pipeline.

When the observations follow an $A-B$ nod pattern, two trace positions can be specified. In this case the pipeline extracts from each frame the spatial profile around each of the two positions, attempts to fit a Gaussian, and uses the goodness of the fit to determine at which position a stellar trace is in fact present. In this way each frame is automatically assigned to the $A$ or $B$ position. If no star trace is found, then it is assumed that that frame is a nod to sky.

When no reference star is available, this step is skipped, and the observations are assumed to be perfectly aligned. Alternatively, the user can manually supply a list of vertical pointing offsets describing the effect of nodding and/or dithering.

Finally, all the observed frames are median-stacked together. When nodding, the $A$ frames and the $B$ frames are stacked independently and then subtracted. This is often very effective in suppressing the strongest sky emission and detector features, and can be used to detect relatively faint emission from the science targets. This stack is created with the sole purpose of allowing a quick check, and is not used in the remaining steps of the data reduction.

\begin{figure*}
  \centering
  \includegraphics[width=0.9\textwidth]{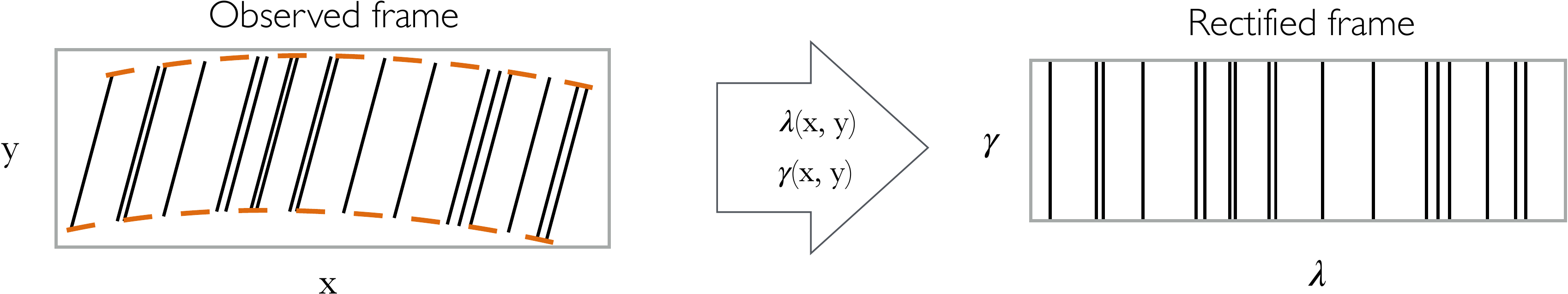}
  \caption{Illustration of the rectification process. The detector registers the data in the \emph{observed frame} $(x,y)$. The 2D function $\lambda(x,y)$, which describes the wavelength solution, is constructed using sky or arc emission lines (shown in black). For the vertical direction, instead, the 2D function $\gamma(x,y)$ is built by tracing the slit edges (shown in orange), and adding a shift derived from the reference star position in the same frame, to account for nodding, dithering, and drifting. These two functions describe a coordinate transformation from the observed frame to the \emph{rectified frame} $(\lambda, \gamma)$. The output grid in $(\lambda, \gamma)$ is arbitrary, and by using the same grid for all frames the data can be stacked without further interpolation.}
   \label{fig:rectification}
\end{figure*}

\subsubsection{Calibrations}
\label{sec:calibrations}

Each of the science frames must be corrected for a number of effects due to cosmic rays, imperfections of the detector, and so on:
\begin{itemize}

  \item Cosmic rays are identified in each frame using L.A.Cosmic \citep{vandokkum01} and masked out (i.e., set to NaN). This is generally not needed when a large number of frames are available, because at the combine stage it is trivial to identify and exclude outlier pixels. However, in some cases the intensity and frequency of cosmic ray hits can create problems for the identification of sky emission lines, and it is best to explicitely mask them out.

  \item The bad pixel mask is constructed by combining pixels in the master flat field that have very low counts (dead pixels) and those in the master dark that have very high counts (hot pixels). Alternatively, it is possible to store and use a default bad pixel mask for a given instrument. Bad pixels are masked out in each frame.

  \item Each science frame is subtracted by the master dark frame and then divided by the master pixel flat field, to account for the slight different sensitivity of each pixel.

  \item The response of detectors to the incoming light intensity is never exactly linear, particularly in the near-infrared, and if not corrected would lead to a biased estimate of relative fluxes. This non-linearity correction can be measured for each detector, and is stored as a set of polynomial coefficients during initialization. This correction is then applied to the science frames.

  \item Finally, the counts are multiplied by the gain and divided by the exposure time to obtain units of electron per second, which will be used throughout the data reduction. For each science frame, a corresponding error frame is constructed by combining in quadrature the read-out noise and the Poisson uncertainty (which is simply the square root of the total counts, in electrons). This error frame is stored as an additional extension to the FITS file.

\end{itemize}

With the exception of the conversion to electron per second and the creation of an error spectrum, all the other steps are optional.

\subsubsection{Slit Identification and Cutout Extraction}
\label{sec:slitid}

In order to properly correct the data for geometric distortion, the slits need to be accurately identified and traced.
A first approximate position of each slit edge on the detector must be stored during the initialization. This can be either calculated from the FITS header or, when the instrument does not store such information, can be manually input by the user.
Starting from these approximate locations, the slit edges are automatically traced on the slit flat field. The user can choose between two methods of slit tracing: either using the continuum emission (ideal for lamp flats and $K$-band sky emission) or using emission lines. In the latter case, the brightest emission lines in the slits are automatically identified, and their top and bottom edges are used to trace the slit. This method works well with the sky emission lines (see, e.g., Figure \ref{fig:example_frame}), and can be applied directly to the science data. Arcs can also be effectively used to trace slit edges.
Each slit edge is then fit using a low-order polynomial as a function of the pixel $x$ coordinate: $\sum e_i x^i$, and the coefficients $e_i$ are saved.

Once all the slits have been identified and traced, they are extracted from each frame and saved as independent FITS files. As these files need to be rectangular, but the slit edges are typically slightly curved, all the pixels outside the traced edges are masked out in the cutouts.

When reducing longslit observations, which typically span a large part of the detector, the user inputs the limits of $y$ pixel coordinates that should be taken in consideration. Then this range is cut out, and treated as the only slit of a multi-slit observation. In this case the slit edges cannot be traced; however, it is possible to use the trace of a bright object in the slit to derive the spatial rectification. If no bright trace is present, then the reduced data will not be spatially rectified, i.e. the trace of an object will not be exactly parallel to the horizontal axis.\\

From this point on, \flame\ operates on one cutout at a time. By cutout here we mean a region of a single frame, therefore each slit will typically have as many cutouts as the total number of observed frames. The running time for the following steps is therefore roughly linear with the number of slits: by setting the data reduction for only one slit the user can obtain a science-quality result for a single object in a short amount of time.

\subsection{Calculation of the coordinate transformations}
\label{sec:transformation}

Typically, spectroscopic data from multi-object instruments present spatial and spectral distortion. The 2D spectra must then be resampled onto a grid that has the horizontal axis parallel to the wavelength direction and the vertical axis parallel to the spatial direction. Furthermore, the grid should be linear in units of wavelengths and spatial pixels. This ``rectification'' is the core of the data reduction, and allows proper sky subtraction and alignment of frames.

We treat the rectification as a coordinate transformation problem. For each individual cutout, the data are obtained in the \emph{observed frame}, described by the $(x,y)$ pixel coordinates. We want to transform the data onto the \emph{rectified frame} of coordinates $(\lambda, \gamma)$, where $\lambda$ is the wavelength, and the vertical coordinate $\gamma$ has units of spatial pixels.

We adopt 2D polynomial functions to describe the coordinate transformation:
\begin{equation}
  \label{eq:lambda_def}
  \lambda(x, y) = \sum_{i,j} \Lambda_{ij} x^i y^j \; ,
\end{equation}
\begin{equation}
    \label{eq:gamma_def}
  \gamma(x, y) = \sum_{i,j} \Gamma_{ij} x^i y^j \; .
\end{equation}

The degree of the polynomials is arbitrary and does not need to be the same for the $\lambda$ and $\gamma$ transformations. Once the matrices of coefficients $\Lambda_{ij}$ and $\Gamma_{ij}$ are known, the data can be resampled on a rectified grid by simply transforming their coordinates, as illustrated in Figure \ref{fig:rectification}. Here we discuss how these coefficients are found.

We note that this approach is similar to the one used, for example, in the data reduction pipelines for the KMOS \citep{davies13} and the MUSE \citep{weilbacher12} instruments at the VLT.

\begin{figure*}
  \centering
  \includegraphics[width=0.8\textwidth]{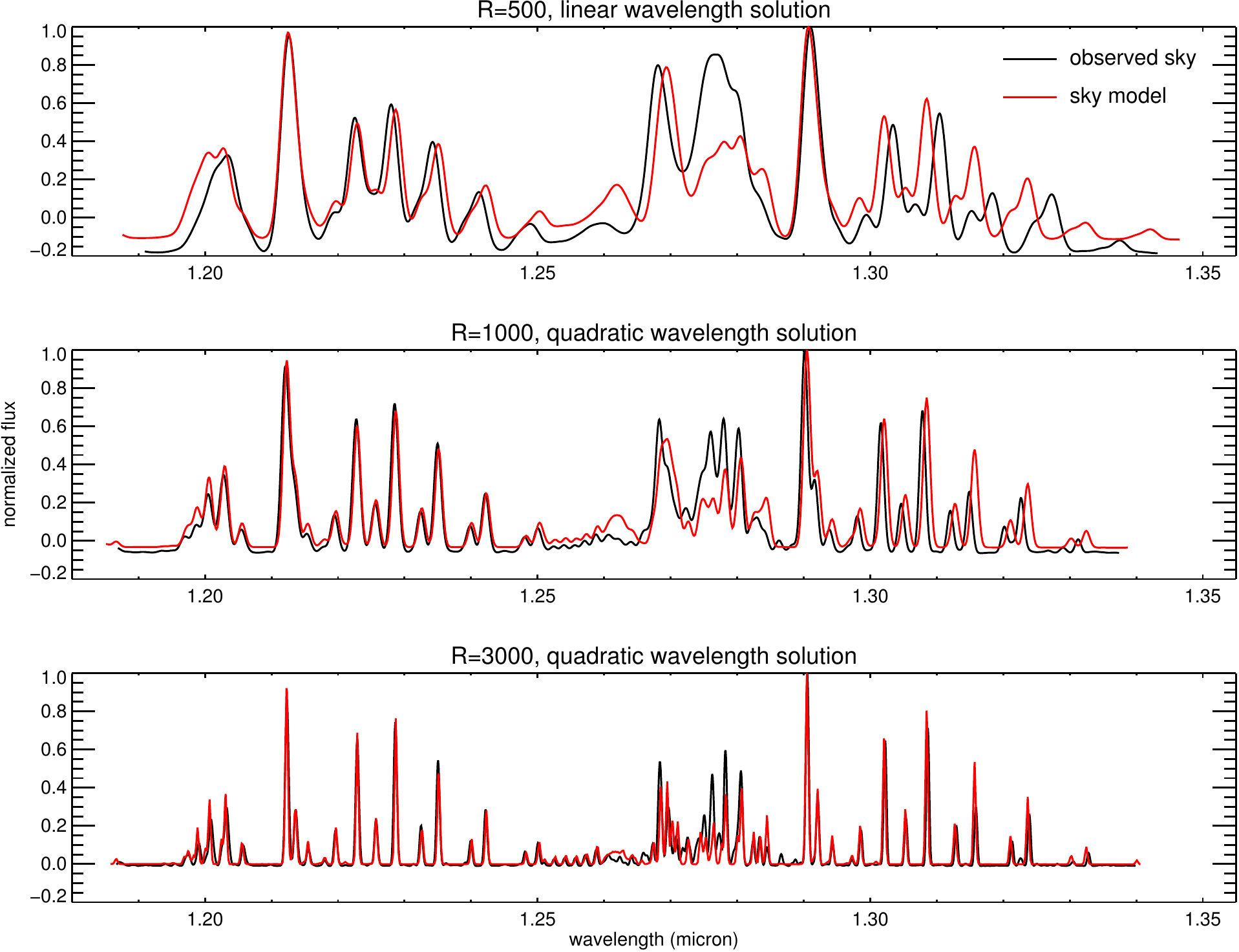}
  \caption{Example of approximate wavelength calibration for LUCI observations in the J band. The observed sky spectrum (black) is compared to a model spectrum (red) via cross-correlation. Initially, the spectra are heavily smoothed, and a coarse grid of dispersion values and initial wavelengths is explored. Then the parameter grids are narrowed, a second-order term is introduced in the wavelength solution, and the fit is repeated at increasingly higher spectral resolution.}
   \label{fig:wavecal_rough}
\end{figure*}

\subsubsection{Spatial Calibration}
\label{sec:spatialcal}

Since the spatial distortion is generally small in multi-slit spectrographs (as opposed to echelle instruments), we can safely assume that the vertical scale is constant throughout the length of the observed frame. With this simplifying assumption, the $\gamma$ coordinate of a pixel is simply its vertical distance from the bottom edge of the slit.

However, with this definition the $\gamma$ coordinate is tied to the slitmask and not to a point in the sky. Nodding, dithering and drifting will cause a shift of the target position with respect to the edge of the slit, and consequently its $\gamma$ value will not be constant. In order to align all frames without further resampling, it is important that $\gamma$ is tied to a fixed point on the sky. To correct for this we add the term $d_\star$ so that $\gamma$ corresponds to the vertical distance in pixels to the reference star:
\begin{equation}
  \label{eq:gamma}
  \gamma(x,y) = y - \sum e_i x^i - d_\star \; .
\end{equation}
The $\gamma$ coordinate therefore follows the true position of the pointing on the sky, and the center of the trace of the reference star corresponds to $\gamma=0$ at all wavelengths.

The coefficients $\Gamma_{ij}$ can be easily derived by comparing Equation \ref{eq:gamma} and Equation \ref{eq:gamma_def}: $\Gamma_{00} = -d_\star - e_0$, $\Gamma_{i0}=-e_i$ (for $i \geq 1$), $\Gamma_{01}=1$, and all the remaining elements are zero. The size of the $\Gamma_{ij}$ matrix is therefore $2\times m$ where $m$ is the degree of the polynomial describing the slit edge (typically, $m=2$ or 3).

While it is generally recommended to choose a spatial pixel scale in the rectified data that is identical to that in the observed data, there are cases when this is not ideal. For example, when observing the same target using the two mirrors of the Large Binocular Telescope, it is possible that the two data sets have slightly different spatial scales. In order to combine the reduced spectra, it is therefore necessary to spatially resample the data. This resampling can be efficiently folded into the spatial calibration by setting $\Gamma_{01} \neq 1$.

%%%%%%%%%%%%%%%%%%%%%%%%%%%%
%\vspace{10mm}%%%%%%%%%%%%%%%  Will not produce a pdf without this
%%%%%%%%%%%%%%%%%%%%%%%%%%%%

\subsubsection{Approximate Wavelength Calibration}
\label{sec:wavecal_rough}

During the initialization, the predicted wavelength range for each slit is calculated. In order to maintain a high level of flexibility, however, \flame\ assumes that the predicted wavelength range is only indicative of the true range. It is therefore necessary to find an approximate 1D wavelength calibration, which has the sole purpose of identifying the rough position on the detector of the spectral features that will later be used for the accurate calibration.

The user can choose whether to use arcs or the sky spectrum for the wavelength calibration. When using the sky spectrum, this is extracted for each slit in the following way. First, the cutout is collapsed along the horizontal direction to obtain the spatial profile. Then, the pixel rows with a flux significantly (3-sigma) higher than the median value are masked, because likely contaminated by bright objects in the slit. Finally, the five most central pixel rows that have not been masked out are taken as representative of the sky, and their median spectrum is extracted. This observed sky spectrum is smoothed to a low spectral resolution, and is compared to a similarly smoothed model spectrum of the sky emission. For the near-infrared sky we use a model based on the ATRAN models \citep{lord92} and released by the Gemini observatory\footnote{\url{http://www.gemini.edu}}, while for the optical sky we use the observed spectrum of \citet{hanuschik03}. The comparison is done via cross-correlation, assuming a grid of pixel scale values that spans the range set during the initialization. The best match yields a rough estimate of the dispersion and the initial wavelength. Despite the fact that the true wavelength solution is almost never linear, this method is robust because is based on matching the overall shape of the smoothed sky spectrum, rather than individual emission lines. Then the procedure is repeated, but a second-order wavelength solution is assumed, and the spectra are smoothed to a slightly higher resolution, so that the brightest emission lines or groups of lines can be used to anchor the solution. Finally, the procedure is repeated a third time, without smoothing the observed spectrum (but the model spectrum is still smoothed to match the instrumental resolution). These steps are illustrated in Figure \ref{fig:wavecal_rough}.

When the wavelength solution is strongly non-linear, the observed spectrum can be optionally split into two halves, and the last step is performed on each segment independently, with a consequently higher degree of freedom in the fit. It is not necessary to find a very precise solution: at the end of this approximate calibration, the wavelength errors need to be smaller than the typical separation between sky lines at the instrumental resolution. For this reason, the approximate wavelength calibration is calculated only for the first frame, and carried over to all the remaining frames.

When arcs are used instead of sky lines, the procedure is slightly different. In this case the emission line wavelengths are known, but their relative brightness is usually not. Also, the calibrations are often taken using a mixture of gases in order to provide a better coverage of the full wavelength range. This means that even if all the line fluxes are known for an individual element, their intensity relative to the lines of a different element is not constrained. For this reason \flame\ treats the arc calibration in a slightly different way. First, the arc spectrum is extracted from the center of the cutout. Then all the bright and isolated emission lines are automatically detected and their centroids are measured via Gaussian fitting. At this point a mock spectrum is constructed by combining a series of Gaussian lines with identical flux and width, placed at the measured pixel positions. This flux-normalized arc spectrum is then compared to a model spectrum constructed in a similar way from the theoretical line lists of the appropriate lamps. The comparison is done in the same way described above for the sky spectrum, and an approximate 1D wavelength calibration for the arc frame is obtained.

\subsubsection{Spectral Line Fitting}
\label{sec:findlines}

Sharp spectral features of known wavelength are needed for an accurate wavelength calibration. Whether arcs or sky emission lines are used, the procedure is the same. The expected wavelengths are stored in a line list, and can be either calculated from first principles or experimentally measured. \flame\ adopts sky emission lines in the optical and near-infrared (with wavelengths measured in vacuum) from \citet{osterbrock96, osterbrock97}, \citet{rousselot00}, and \citet{hanuschik03}. To avoid blended lines, three different line lists are available for different spectral resolution ($R\sim1000$, 3000, and 6000). During the initialization the appropriate line list is selected and a local copy is saved in the working directory so that, if necessary, lines can be easily added or removed by the user.

First, the spectrum of the central region of the slit is extracted and, adopting the approximate wavelength calibration found in the previous step, the emission lines from the line list are identified. The position of every emission line that has been identified is measured via Gaussian fitting. This yields a list of measured $x$ positions and corresponding expected wavelengths, and a polynomial wavelength solution is robustly fit to these measurements. Using this wavelength solution, more lines are identified and a more accurate solution is found, in an iterative way. The loop ends when no new emission lines are identified.

Next, the emission lines that were identified in the central region of the slit must be traced across the slit. This is done by extracting the spectrum of each pixel row and using the previous wavelength solution to identify the lines. As long as the wavelength solution varies smoothly with the position along the slit, the result of the previous pixel row can be used as an accurate guess. Starting from the central pixel row, the solution is propagated first to all the pixel rows in the top half, and then to those in the bottom half. At each step the previous 1D wavelength solution is discarded; its only purpose is to identify the emission lines at that pixel row.

In order to obtain the most accurate wavelength calibration, spectral features of unknown wavelengths are also traced and used to improve the rectification, as explained below. These features are included in the line list with a flag indicating that their wavelength should be considered as an approximate value. At each pixel row, these features are identified and measured together with the emission lines of known wavelength, but they are not used to find the temporary 1D wavelength solution.

\begin{figure}
  \centering
  \includegraphics[width=\columnwidth]{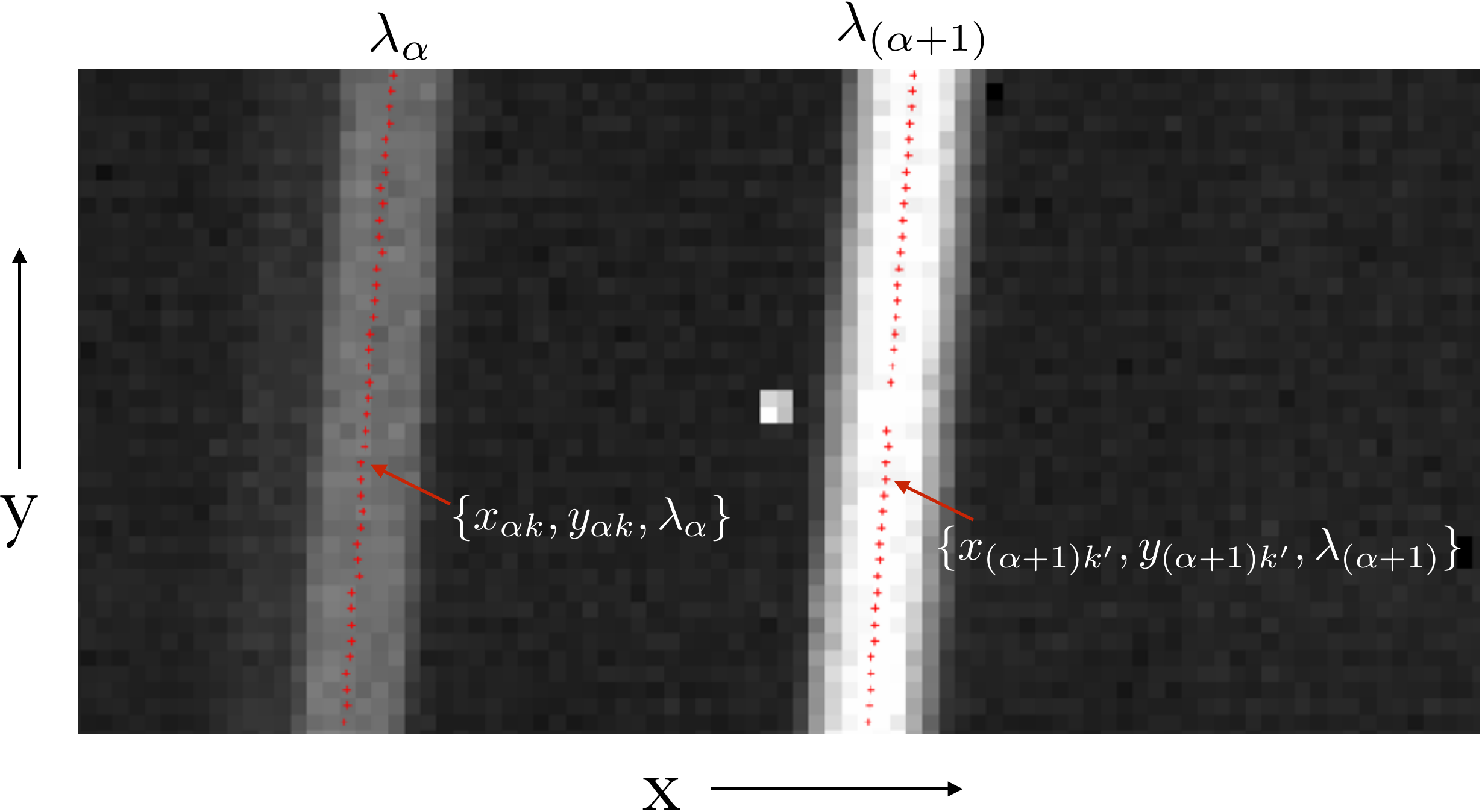}
  \caption{Sky or arcs emission lines are identified and their positions measured at each pixel row. Each detection, marked by a red cross, consists of three quantities: the measured $x$ pixel coordinate, the $y$ coordinate of the pixel row, and the wavelength $\lambda$ of the emission line. The number of detections can be different for different emission lines, because not every pixel row yields a reliable measurement. In the LUCI observations shown here as example, a cosmic ray falling near an emission line compromised the line detection in two pixel rows.}
   \label{fig:two_lines}
\end{figure}

The final result is a set of emission line detections: $ \{ x_{\alpha k}, y_{\alpha k}, \lambda_\alpha \} $ where $\alpha$ is the index that identifies the spectral line and $k$ runs through the pixel rows. This is illustrated in Figure \ref{fig:two_lines}. Not all pixel rows are necessarily present in the final list of identification, because fits can fail due to poor signal, bad pixels, or cosmic rays.

\subsubsection{Wavelength Calibration}
\label{sec:wavecal}

The wavelength calibration, i.e., the function $\lambda(x,y)$ that describes the wavelength at each pixel in the observed frame, can be easily obtained by fitting a smooth surface to the measured points $ \{ x_{\alpha k}, y_{\alpha k}, \lambda_\alpha \} $. In practice, this means adopting Equation \ref{eq:lambda_def} to describe the wavelength solution, and finding the coefficients $\Lambda_{ij}$ that minimize the wavelength residuals:
\begin{equation}
  \mathrm{Res}(\Lambda_{ij}) = \sum_\alpha \sum_k \left( \sum_{ij} \Lambda_{ij} x_{\alpha k}^i y_{\alpha k}^j - \lambda_\alpha \right)^2 \; .
\end{equation}
The least-square minimization is performed using \texttt{mpfit} \citep{markwardt09}. For narrow slits in singly-dispersed multi-object spectrographs, a good description of the wavelength solution can be typically obtained using a polynomial of third degree along $x$ and second degree along $y$ (i.e., $\Lambda_{ij}$ is a $4 \times 3$ matrix).

This technique, which is the standard method to derive the wavelength solution, can be used with either sky or arc lines, and works well if the emission lines of known wavelength are isolated and cover the majority of the slit in a homogeneous way. However, blended lines cannot be used, because even if the wavelength of the individual lines are known, the effective wavelength of the blend is typically unknown, and can even change with time due to varying line ratios. Moreover, lines in arcs or sky spectra for which a precise wavelength is not available must also be discarded. While in most cases these do not represent serious issues, there are specific situations (e.g., in the $K$ band, or near the 1.27 $\mu$m oxygen band) where the lack of reliable, isolated lines sets a limit on the accuracy of the wavelength solution. This causes two problems: first, the absolute wavelength scale of the final spectra will not be accurate; second, the error in the relative wavelength solution of different pixel rows across the slit will produce ``wavy'' sky lines in the rectified frame, which cannot be reliably modeled or subtracted.

In order to make use of all the information in the data, we include spectral features of unknown wavelength in the wavelength calibration. They can be isolated lines, blends, or complex features, as long as their shape is constant across the slit (but it can vary from frame to frame). The idea is to use the fact that a given feature must be at the same wavelength in all pixel rows, which is a useful constraint on the 2D wavelength solution, despite the fact that the wavelength is not known. Instead of the expected wavelength, for these features we use the median observed wavelength in all pixel rows:
\begin{equation}
  \lambda_\alpha =
    \begin{cases}
      \lambda_{\text{th}, \alpha} & \text{if known} \\
      \underset{k}{\mathrm{median}} \left( \sum_{ij} \Lambda_{ij} x_{\alpha k}^i y_{\alpha k}^j \right) & \text{otherwise}
    \end{cases}
\end{equation}
where $\lambda_{\text{th}, \alpha}$ is the theoretical (in the sense of previously known) wavelength of the emission line $\alpha$.

When possible, the wavelength calibration should be derived from sky emission lines rather than arcs, because the sky spectrum is present in the science observations and allows an accurate calibration for each frame. When using arcs, on the other hand, one needs to worry about slight changes in the wavelength calibration between the science and the arc lamp observations, and also among different science frames, especially for long exposure. For this reason when deriving the wavelength calibration using arcs, the sky spectrum is extracted from each science frame and the brightest emission lines are used to shift the wavelength solution. For very blue spectra (bluer than [\ion{O}{i}]5577) no bright emission lines are available; in these cases the sky continuum is extracted and the wavelength shift is derived by cross-correlating the observed sky spectrum to a model spectrum. However this technique is not as reliable, due to the very faint sky emission at blue wavelengths. It is advisable, in these cases, to turn off the automatic wavelength shift and take the calibrations carefully, for example making sure that the arcs are taken at the same elevation as the science data.

The wavelength calibration is a critical step of the data reduction, and determines whether the spectra can be successfully rectified and sky-subtracted. Figures and tables are generated by \flame\ to help the user assess the quality of the wavelength solution. Users can iteratively change the settings (for example, the polynomial degrees of the wavelength solution, or the sigma-clipping threshold used to reject outliers during the fit) until they reach the desired precision. Most times it is possible to obtain a good wavelength calibration, with a median absolute deviation of the wavelength residuals smaller than a tenth of a pixel, just by removing from the line list those lines that have not been properly identified or traced.

\subsection{Operations on Slits}

After the coordinate transformations from the observed to the rectified frame have been found, data reduction steps on the individual cutouts can be performed. These consists of illumination correction, sky subtraction, rectification, stacking, testing of the quality of the data reduction, and extraction of 1D spectra.

\subsubsection{Illumination Correction}
\label{sec:illumcorr}

In this optional step, each cutout is corrected for the uneven illumination of the slit along the spatial direction using either the sky emission lines or the illumination flat field.

If sky lines are being used to derive the wavelength calibration, then they will also be used to measure the illumination function. When Gaussian fits are performed on each of the emission lines as described in Section \ref{sec:findlines}, in addition to the position and wavelength, the flux $f_{\alpha k}$ is also measured. Every sky line has a different intensity, so for each line we divide the flux measurements by their median values: $ f_{\alpha k} / \mathrm{median}_k(f_{\alpha k}) $. These normalized fluxes contain the information on the relative illumination along the slit. For each emission line fit we also calculate the rectified spatial coordinate: $\gamma_{\alpha k} = \gamma(x_{\alpha k}, y_{\alpha k})$. The illumination function $I(\gamma)$ is then constructed by fitting all the normalized flux measurements as a function of $\gamma$.

When sky lines are not sufficient for the wavelength calibration, the arcs can also be used to derive the illumination function. However, the illumination function derived using light emitted by lamps in the dome may not match the response of the spectrograph to the light coming from distant targets. A better approach is to use illumination flats, possibly obtained by observing the sky in twilight. In this case the median spectrum of the sky is divided out and the average illumination function $I(\gamma)$ is obtained.

Once the illumination function is known, the data are corrected directly in the observed frame, i.e. for each pixel $(x,y)$ in the cutout the flux is divided by $I(\gamma(x,y))$.

\begin{figure}
  \centering
  \includegraphics[width=\columnwidth]{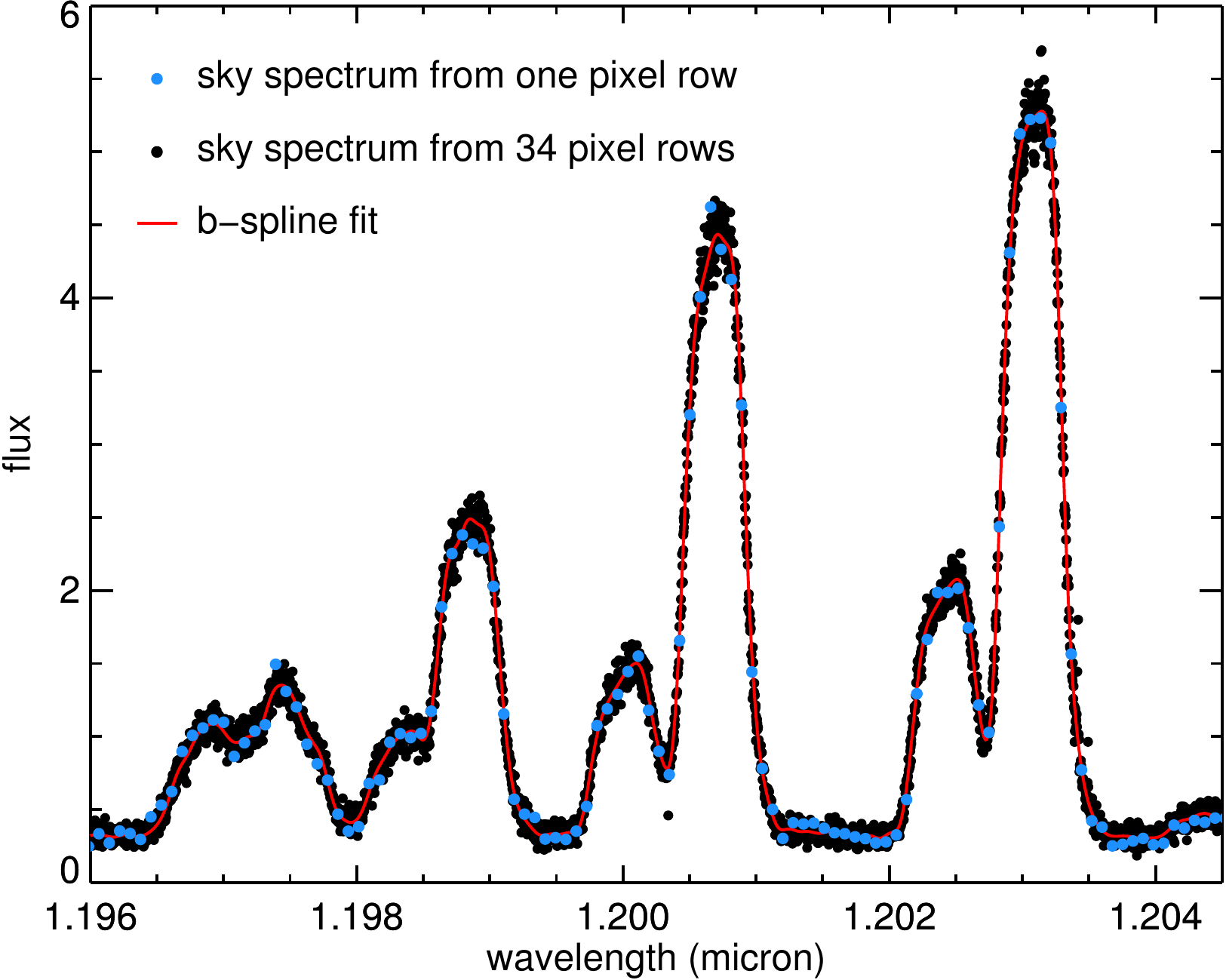}
  \caption{Example of sky subtraction for LUCI observations in the J band. In this case the slit is empty and is 34 pixel tall, which means that using the sky emission from the full frame (shown as black points) gives a spectral sampling that is more than one order of magnitude finer than what can be obtained using only one pixel row (shown as blue points). The 1D model of the sky (shown in red) is then fit to the data using b-splines.}
   \label{fig:skysub}
\end{figure}

\subsubsection{Sky subtraction}
\label{sec:skysub}

One of the main goals of any data reduction pipeline is to subtract as accurately as possible the sky emission from the data. There are two different ways to achieve this: one is to observe a sky frame and subtract it from the science frame (nodding); the other is to build a model of the sky spectrum starting from the science data. Both methods are implemented in \flame\ and can be used together. In section \ref{sec:combine} we will discuss how nodding is handled during the data reduction, while here we explain how the sky model is built and subtracted from the data, following the method described by \citet{kelson03}.

The key idea is to take advantage of the distortion in the 2D wavelength solution to better sample the sky spectrum. As shown in Figure \ref{fig:two_lines}, the emission lines are typically tilted with respect to the detector vertical axis. This is due either to the fact that the slit was tilted on purpose, or to geometric distortions. Since the wavelength of each pixel in the observed frame is known, we can extract the spectrum of the sky using all the pixels in the slit that are not contaminated by bright targets, rather than extracting the spectrum from a single pixel row. The substantial oversampling offered by the full observed frame allows one to construct a 1D model of the sky spectrum with a high accuracy. This is illustrated in Figure \ref{fig:skysub}, where the blue points show the sky spectrum extracted from the central pixel row of a slit, and the black points show what can be obtained when using the full slit.

The sky spectrum is modeled using a one-dimensional b-spline function. By default the breakpoints are set equal to the pixel positions in the central pixel row of the frames. By changing the distance between successive breakpoints, the user can choose the degree of flexibility of the b-spline. The ideal spacing of the breakpoints is different for each observation and depends on the instrumental resolution, slit height, slit tilt, and so on. For example, with a uniform and frequent sampling, such as that shown in Figure \ref{fig:skysub}, a denser set of breakpoints can be used for a more accurate description of the sky spectrum at the sub-pixel level.

It is critical to exclude pixels that are affected by strong emission from astronomical sources, otherwise the sky model will be biased and the scientifically interesting emission may be over-subtracted. To avoid this, an iterative outlier rejection is applied during the b-spline fitting. At each step, a best-fit 1D model of the sky is fit to the flux values as a function of their wavelengths, and the pixels with the largest absolute deviation from the best-fit model within a running window are discarded. The fraction of discarded pixels, the size of the window, and the number of iterations can be set by the user. Being able to fine-tune these parameters can be useful in cases where the object covers a substantial fraction of the slit and needs to be carefully filtered out.

Fitting a 1D model of the sky emission relies on the assumption that the sky spectrum does not vary along the slit. While this is generally true, imperfections in the slitmask and in the illumination may introduce spatial variations in the sky spectrum. In these cases it is important to apply a good illumination correction, as described above.

Once the 1D sky model $s(\lambda)$ is obtained as a function of wavelength, a 2D version is generated in the observed frame by using the coordinate transformations: $S(x,y)=s(\lambda(x,y))$. The 2D sky model is then subtracted from the data. In \flame\ the final output is always produced both with and without sky subtraction, and the user is free to select the result that is more appropriate.

\subsubsection{Rectification}
\label{sec:rectification}

Before finally stacking the frames, it is necessary to resample the data from the observed frame $(x,y)$ to the rectified frame of coordinates $(\lambda, \gamma)$. This transformation is mathematically defined by the functions $\lambda(x,y)$ and $\gamma(x,y)$. However, we still need to define the pixel grid in the rectified frame, which is in principle arbitrary. In order to avoid heavy under- or over-sampling, it is best to have a similar pixel spacing in the observed and rectified frames. A good choice of the output grid is also critical for reducing the number of interpolation steps, as we explain below.

The wavelength grid is set in the following way: the median wavelength pixel scale is measured from the wavelength-calibrated first frame, and is then rounded to the closest number on a relatively coarse logarithmic scale. This ensures that data with a slightly different wavelength solution will have identical pixel scale. Then, the wavelength of the first pixel of the rectified grid is chosen as an integer number of pixels from a fixed reference wavelength, which is arbitrarily set to 1 $\mu$m. This means that rectified spectra observed at different wavelengths lie on the same regular grid, and can be stacked or merged without having to perform an additional interpolation.

For the spatial grid we use a different approach. The pixel scale is always set to one, so that the rectified and the observed frames have the same spatial scale (unless the user has set $\Gamma_{01} \neq 1$, as discussed in Section \ref{sec:spatialcal}). The $\gamma$ value of the first and last pixels are then set to the minimum and maximum $\gamma$ values measured in the observed frame, rounded to the nearest integer. This means that successive frames can have substantially different spatial grids, particularly when nodding. Despite this, the data can be stacked without interpolation, because a shift by an integer number of pixels is sufficient to match different grids.

Once the output grid has been defined, it is necessary to resample the observed data. This is the only step in \flame\ where the data are interpolated. The default interpolation method used by \flame\ is a natural neighbor scheme: the data value at an arbitrary point is given by the linear combination of the values in the neighboring pixels, with weights derived via Voronoi tassellation. Other options, such as nearest neighbor, linear, and quintic interpolation are available, offering different degrees of compromise between speed and performance \citep[for a discussion on the interpolation methods in the context of rectification, see][]{davies13}.

The rectification is performed identically on the data, the sky-subtracted data, the 2D model of the sky, and the corresponding uncertainties.

\subsubsection{Combining the frames}
\label{sec:combine}

For each slit, the rectified cutouts corresponding to different frames must be stacked together. We first consider the case of observations that are taken without nodding.

By construction, all rectified cutouts have the same $\lambda$ grid (i.e., same wavelength pixel scale, wavelength range, and number of pixels along the horizontal direction). However, if drifting or dithering was present at the level of one pixel or more, the $\gamma$ grid may be slightly different for different frames. As explained in Section \ref{sec:rectification}, these grids differ by an integer number of pixels, so that the cutouts can be aligned along the $\gamma$ axis by simple vertical shifts, without the need for interpolation. As a result, the output grid will be identical to the input ones along the $\lambda$ direction, but may be slightly wider along the $\gamma$ direction, in order to accommodate all the pixels from the individual cutouts.

The frames are combined with a weighted mean, and the user can choose what type of weight $W_i$ to use for each frame. The default option, and the only one available when no reference star is used, is to have identical weights (i.e., the frames are combined with an arithmetic mean). Otherwise, it is possible to use as weight the flux of the reference star trace, calculated for each frame as described in Section \ref{sec:diagnostics}. This choice maximizes the contribution of those frames with clearer conditions. The weight can also be set equal to the inverse of the FWHM of the reference star trace, and in this case the spatial resolution is maximized. Finally, the Gaussian peak of the reference star trace can also be chosen as weight, which is a compromise between transmission and spatial resolution.

To filter out cosmic rays, at each pixel position the flux values are sigma-clipped before being stacked. The clipping is done around the median, and is typically very effective in rejecting cosmic ray hits. We call $m_i(\lambda, \gamma)$ the mask for the $i-$th frame, which is 1 for the good pixels and zero for the pixels that must be discarded, i.e. either cosmic ray detections or bad pixels.
By combining the frame weights with the bad pixel mask we can construct a weight map for each cutout: $w_i(\lambda, \gamma) = W_i \cdot m_i(\lambda, \gamma)$. If the flux in the rectified cutout is $f_i(\lambda, \gamma)$, we can then write the stacked spectrum as:
\begin{equation}
  \label{eq:stack_flux}
  F(\lambda, \gamma) = \frac{\sum_i \, w_i(\lambda, \gamma) \, f_i(\lambda, \gamma)}{\sum_i w_i(\lambda, \gamma) } \; .
\end{equation}

As explained in Section \ref{sec:calibrations}, for each science frame a corresponding error frame is generated by combining the Poisson and the read-out noise. These error frames are carried throughout the data reduction, and undergo the same rectification applied to the science frames. When frames are stacked, the corresponding error frames are stacked in quadrature, yielding the final error frame:
\begin{equation}
  \label{eq:stack_err_th}
  \sigma_\text{th}(\lambda, \gamma) = \frac{ \sqrt{ \sum_i \left[ w_i(\lambda, \gamma) \, \sigma_i(\lambda, \gamma) \right]^2 } }{ \sum_i \, w_i(\lambda, \gamma)} \; .
\end{equation}
In addition to this \emph{theoretical} noise, \flame\ calculates also an \emph{empirical} error spectrum. This is based on the variation of the pixel value in different rectified frames, and is defined as:
\begin{equation}
  \label{eq:stack_err_emp}
 \sigma_\mathrm{emp}(\lambda, \gamma) = 1.5 \cdot \sqrt{ \frac{ \sum_i w_i^2 }{ \left(\sum_i w_i \right)^2 - \sum_i w_i^2 } \cdot \frac{ \sum_i w_i (f_i-F)^2 }{ \sum_i w_i } }  \; ,
\end{equation}
where for simplicity we dropped the arguments $(\lambda, \gamma)$ for the weights and the flux values. When all weights are identical, this formula reduces to the standard error of the mean $1.5 \sqrt{\text{Var}(f_i)/N}$, where the factor 1.5 is needed to account for the correlated noise due to the fact that the rectified frames have been resampled. The theoretical explanation for this factor is given by \citet{fruchter02} and relies on a number of assumptions; however, \citet{kriek15} obtain the same result by analyzing how random resampling affects the noise properties of real spectroscopic data. This empirical estimate $\sigma_\mathrm{emp}$ is clearly more accurate when the number of frames is large. In Section \ref{sec:performance} we will show the comparison of the theoretical and the empirical error spectrum for example data sets.

The two types of error spectra are saved as additional extensions to the FITS file containing the final stack. The 2D model of the sky emission, the map of the effective exposure time for each pixel (i.e., excluding rejected frames), and the weight map are also added as FITS extensions.

When observations are taken with an $A-B$ nodding, the stacking described above is carried out independently for the $A$ frames and for the $B$ frames, which are then subtracted in order to obtain the $A-B$ stack. This method is different from the more common approach of associating each frame to a sky frame, and then stacking individual $A-B$ pairs \citep[e.g.][]{kriek15}. Our approach works best when applied to frames that have already been sky subtracted. The advantage is that the observations do not need to be an exact sequence of alternating $A$ and $B$ nod positions. This allows more flexibility while observing and also when discarding frames because of poor conditions. However, the quality of the resulting sky subtraction will of course depend on the sampling of the sky lines: a regular and frequent sampling will yield the best results.

If the nodding is off-slit (see Figure \ref{fig:nodding_diagram}), then the $A-B$ stack represents the final output. For on-slit nodding, a further step is needed in order to combine the positive signal from the $A$ frames with the negative signal from the $B$ frames. This is done by adding the $A-B$ stack to the $B-A$ stack, after this has been shifted in order to align the $\gamma$ coordinates. If nodding is performed on paired slits instead, then the outputs of the two slits need to be combined together. Paired slits are automatically identified by comparing the nodding length to the vertical distance between slit centers.

Finally, all the slits that were reduced are pieced together in two large FITS files, one containing the flux and the other the signal-to-noise ratio, which is particularly useful for detecting faint emission lines. These files are not meant to be used in the scientific analysis, but only as an easy way to explore the final reduction of all the slits.

All the steps described above are performed twice: once using the rectified frames and once using the frames that have been sky-subtracted (as explained in Section \ref{sec:skysub}) and rectified. The outputs are organized into two subdirectories that contain the same number and type of files, and the user is free to choose which version of the final output to adopt. Generally, the sky-subtracted output is better, but there are cases in which the sky modelling does not yield good results, for example when the emission from the target spans most of the slit along the spatial direction.

When one target (or slitmask) is observed over multiple nights, the different sets should be independently reduced using the appropriate calibrations, and then combined. The \flame\ code includes a stand-alone routine that can be used to combine the output from different data reduction runs, preserving the data structure of the FITS files and correctly combining also the secondary data products (i.e., the uncertainty, sky, exposure, and weight maps).

\begin{figure}
  \centering
  \includegraphics[width=\columnwidth]{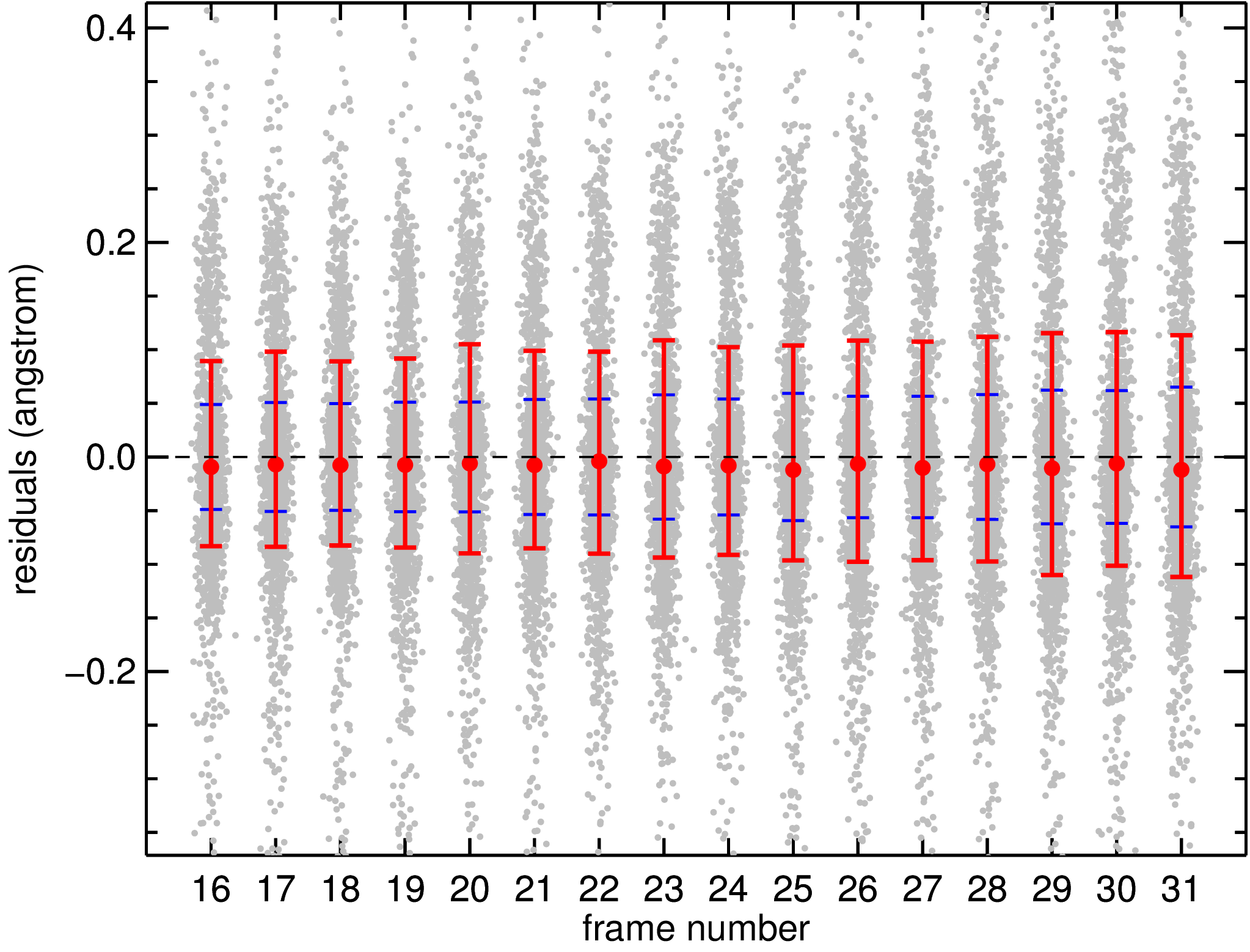}
  \caption{Example of quality check for a slit observed in the $J$ band with LUCI. Each gray point represents the wavelength residual measured for a sky line in a single pixel row in an individual cutout. The blue bars show the median absolute deviation, while the red points and error bars mark the median values and the central 68th percentile.}
   \label{fig:datacheck}
\end{figure}

\subsubsection{Quality check}
\label{sec:datacheck}

An automatic quality check is run on the reduced data. While some of the previous steps output informative plots on the wavelength calibration and rectification of individual frames, the quality of the overall reduction can only be assessed by analyzing the final stacked and rectified data.

First, the slit containing the reference star is analyzed. The median spatial profile of the star is extracted, and its FWHM is measured via Gaussian fitting. This quantity, converted into arcseconds, gives a useful measure of the effective seeing throughout the observations. The FWHM and vertical position of the stellar trace are also measured and plotted as a function of wavelength, in order to verify the consistency of image quality and spatial rectification.

Then, the quality of the wavelength solution and rectification is tested for each slit. Sky lines are identified and measured from the rectified and stacked sky model, and their wavelength residuals and line widths are plotted as a function of wavelength. These measurements are also used to derive the effective spectral resolution and the uncertainty in the wavelength solution.

\begin{figure*}
  \centering
  \includegraphics[width=\textwidth]{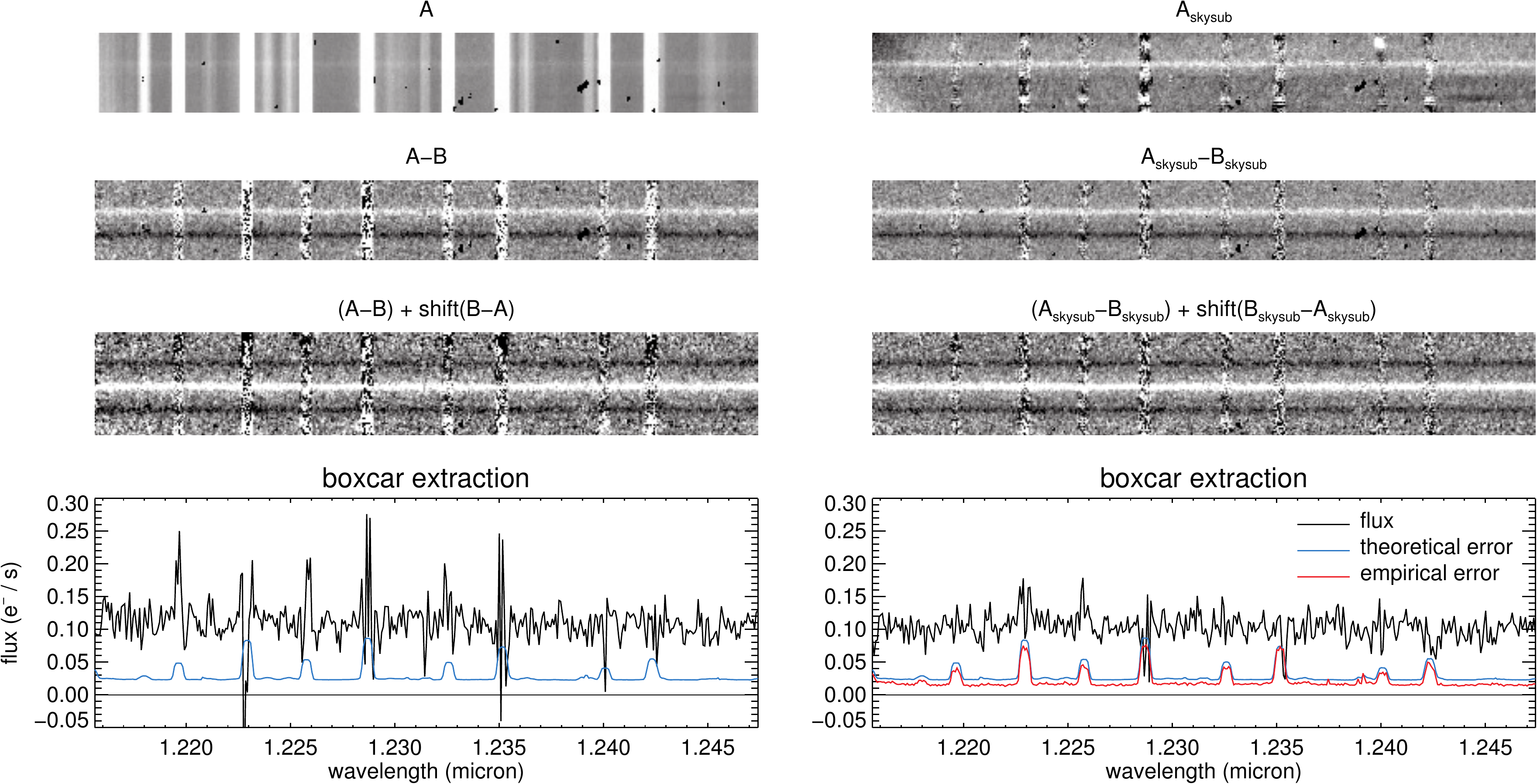}
  \caption{Example of LUCI data reduction. The target is a high-redshift galaxy with a smooth, featureless continuum emission. The two columns show successive steps in the stacking of the frames, without (left) and with (right) modelling and subtracting the sky emission in individual frames. Only a part of the full 2D spectrum is shown. Top row: stack of $A$ frames. Second row: difference between the stack of $A$ frames and the stack of $B$ frames. Third row: sum of $A-B$ and $B-A$ stacks, after vertical alignment. Bottom row: boxcar extraction centered on the continuum emission of the target. The theoretical (blue) and empirical (red) error spectra are also shown. If the frames are not individually sky-subtracted, the frame-to-frame variation is dominated by the physical variation of the OH line flux, therefore the empirical error cannot be calculated.}
   \label{fig:example_LUCI}
\end{figure*}

When the wavelength solution is based on sky lines rather than arcs, it is possible to check the quality of the rectification in individual frames by looking at the wavelength residuals for each of the measured sky lines. An example of such diagnostic plot is shown in Figure \ref{fig:datacheck}, where the distribution of wavelength residuals as a function of frame number is shown. This is a very robust way to determine the quality of the wavelength calibration, and can be used for a thorough frame selection on a slit-by-slit basis. Other plots that are produced include the distribution of line widths, to check the quality of focus and spectral resolution over time; and the distribution of residuals for each of the sky lines, to identify lines that are systematically offset from the wavelength solution, which may be due either to a line misidentification or to an inadequate polynomial order for the function $\lambda(x,y)$.

\subsubsection{Extraction}
\label{sec:extract}

The final step consists of the extraction of 1D spectra from the reduced and combined frames. For each slit, the spatial profile is constructed and fit with a Gaussian curve. If the fit yields a signal-to-noise ratio larger than 5, then the spectrum is extracted; otherwise, no extraction is performed. This test is carried out independently on both the output sets (with and without sky subtraction) and the results are saved into two different subdirectories.

By default \flame\ performs an optimal extraction \citep{horne86} using the Gaussian fit to the spatial profile as weight, but the user can change these settings. Plots showing the extracted profile and the Gaussian fit are produced, and can be used to check that the correct object has been extracted. Additionally, a stand-alone extraction routine is provided, with which users can obtain a boxcar extraction (with arbitrary aperture) from any reduced frame, including those for which no trace could be identified.

Each 1D spectrum is saved in a FITS file as a structure containing the following fields: wavelength, flux, empirical and theoretical uncertainty, and sky emission. The format of these files is compatible both with SpecPro\footnote{\url{http://specpro.caltech.edu}} \citep{masters11} and SpecViz\footnote{\url{https://github.com/spacetelescope/specviz}}, which are two publicly available interacting programs for viewing 1D spectra.

% -------------------------------------------------
%  					PERFORMANCE AND EXAMPLES
% -------------------------------------------------

\section{Performance and Examples}
\label{sec:performance}

We illustrate the performance of \flame\ by showing two examples of data reduction, one in the near-infrared and one in the optical.

\subsection{Example of Near-Infrared Data Reduction: LBT/LUCI}

LUCI is a twin multi-object near-infrared spectrograph at the Large Binocular Telescope \citep{seifert03}. Two almost identical spectrographs, LUCI1 and LUCI2, are mounted on each of the two 8.4-meter mirrors, and can be used simultaneously, with either identical settings or with different sets of filters, grating angles, etc.

We test \flame\ by reducing a set of 39 frames of 5 minutes, corresponding to a total of 3 hours and 15 minutes of exposure time. The data were taken with LUCI1 using a slitmask with one slit on a bright reference star and six slits centered on faint distant galaxies. The observations are in the $J$ band, with the G210 grating and the N1.8 camera. The $1\arcsec$ slit width yields a spectral resolution $R\sim 3000$. The observations were taken with an on-slit nodding pattern, with the $A$ and the $B$ positions offset by $3\farcs5$. Additionally, a dithering of $\pm 0\farcs25$ was applied to each of the two nod positions (i.e., the pattern was $ABA'B'$).

We run the pipeline with the default settings for LUCI. Slit edges are traced using the sky OH lines in the science frames. Sky lines are also used for the wavelength solution (so that arcs are not needed) and for the illumination correction. No explicit cosmic ray rejection is needed, because bright, long traces that can heavily affect the wavelength calibration are typically not a problem with near-infrared detectors, particularly when the exposure times are short.

Using a Linux machine with 8 GB of RAM and a 3.4 GHz Intel Core i5 processor, the data reduction of the 39 frames took 40 minutes. The complete reduction of just one of the seven slits took 10 minutes.

The final combined data show wavelength residuals with a typical median absolute deviation of 0.05\AA, corresponding to about 3 km s$^{-1}$.

To illustrate the quality of the sky subtraction, we consider a small portion of one of the seven slits, targeting a high-redshift galaxy. In the top panels of Figure \ref{fig:example_LUCI} we show the stack of the $A$ frames without (left) and with (right) subtraction of a model sky spectrum from individual frames. The faint continuum emission from the galaxy is barely detected in the first panel, because hidden by the bright sky emission. The sky subtraction is clearly effective in making the galaxy spectrum more visible; however it also leaves strong residuals from the sky lines, and numerous detector defects. These issues can be improved by using a pixel flat field, which can remove the difference in pixel sensitivity that causes some of the systematics. While it is possible to obtain a good pixel flat for LUCI by using halogen lamps, the sensitivity pattern vary substantially over time, and we found that using a pixel flat that was not taken together with the science observations often has the opposite effect of enhancing these patterns.

Fortunately, these residuals are cancelled by the $A-B$ nodding, as is clear from the panels in the second row of Figure \ref{fig:example_LUCI}. Here the $A-B$ difference is shown, again without and with sky subtraction applied to the individual frames. Clearly, nodding is extremely effective in removing sky residuals and detector issues. Without explicit sky subtraction on the individual frames, however, the sky lines are still present, because their variable flux is not perfectly subtracted out. In this example, the positive residuals indicate stronger emission lines in the $A$ frames. Clearly, the sky-subtracted stack is less noisy, without visible systematics in the regions affected by sky emission. The higher noise present in the regions interested by sky emission is for the most part due to the unavoidable Poisson shot noise, as we demonstrate below.

The third row of Figure \ref{fig:example_LUCI} shows the result of combining the $A-B$ stack with the $B-A$ stack, after a shift equal to the nod amplitude. In these frames, which represent the final output of \flame, the central positive trace of the object is the sum of the signal in all the $A$ and $B$ frames, while each of the negative traces contains only half of the signal. The version that is not sky subtracted shows strong residuals at the top and bottom of the slit, where only one of the two stacks is present. However, in the central part the positive sky residuals in the $A-B$ frames are canceled by the sky residuals in the $B-A$ frames, which are equal in intensity but opposite in sign. These residuals in the central part of the slit look qualitatively similar to the residuals in the sky-subtracted version of the same frame.

Finally, the bottom panels show a simple boxcar extraction of the 1D spectrum of the galaxy from the final stacks. The corresponding error spectrum, plotted in blue, shows clearly the enhanced noise corresponding to the sky emission lines. Despite the partial removal of the sky residuals in the final combine, the stack without sky subtraction still presents significant residuals due to OH lines, whereas in the sky-subtracted spectrum the residuals are mostly consistent with the predicted error. We thus conclude that the best approach is to first model and subtract the sky in each cutout, and then apply the $A-B$ subtraction \citep[see also][]{steidel14,chilingarian15}.
For the sky-subtracted stack, it is also possible to calculate the empirical error given by Equation \ref{eq:stack_err_emp}, shown as a red line in Figure \ref{fig:example_LUCI}, which is in excellent agreement with the theoretical estimate.
Since the theoretical error is entirely determined by the combination of readout and Poisson noise, we conclude that the sky subtraction in \flame\ is able to reduce the noise at the location of sky lines to levels that are comparable to the Poisson limit.
The empirical estimate of the error cannot be calculated for the non-sky-subtracted stack, because in that case the frame-to-frame variation is mainly due to a physical variation of the sky line emission rather than to statistical fluctuations.

\subsection{Example of Optical Data Reduction: Keck/LRIS-BLUE}

LRIS \citep{oke95} is a multi-object spectrograph mounted at the Keck I telescope, and features a red and a blue arm. The red arm \citep{rockosi10} probes the wavelength range 6000~\AA\ to 1 $\mu$m, a spectral region where the OH sky lines are relatively strong. For this reason many techniques shown above for near-infrared data (such as wavelength calibration and illumination correction) can be applied to observations taken with the LRIS red arm, even though these data are obtained with red-sensitive CCDs rather than near-infrared detectors. The main differences are that nodding is typically not needed (partly because CCDs are less affected by defects compared to near-infrared detectors), but cosmic ray hits are strong and numerous, so that cleaning of individual frames using L.A.Cosmic is advisable.

The blue channel \citep{mccarthy98, steidel04}, on the other hand, is sensitive to the wavelength range 3000 - 6000~\AA\ and requires slightly different methods. In order to test \flame\ with optical observations we downloaded from the Keck Observatory Archive\footnote{\url{https://koa.ipac.caltech.edu}} a sample data set (PI: C. C. Steidel) obtained with the LRIS blue arm. The observations were taken using the 600/4000 grism, corresponding to a spectral resolution $R\sim1000$ for a slit width of $1\arcsec$. Five frames of 30 min each were observed without nodding nor dithering, for a total of 2hr 30min on source. The multi-slit mask contains 31 slits targeting faint high-redshift galaxies.

Since the LRIS raw data do not conform to the \flame\ requirements, the initialization routine converts them into the standard format. First, raw frames consist of different FITS extensions corresponding to different amplifiers, which must be read and saved as a single, contiguous FITS frame. Second, each amplifier has a slightly different value for the gain; these values are applied at this step, and the overall gain is set to 1. Finally, since the wavelength runs along the vertical direction, the frames need to be rotated by 90 degrees.

\begin{figure}
  \centering
  \includegraphics[width=\columnwidth]{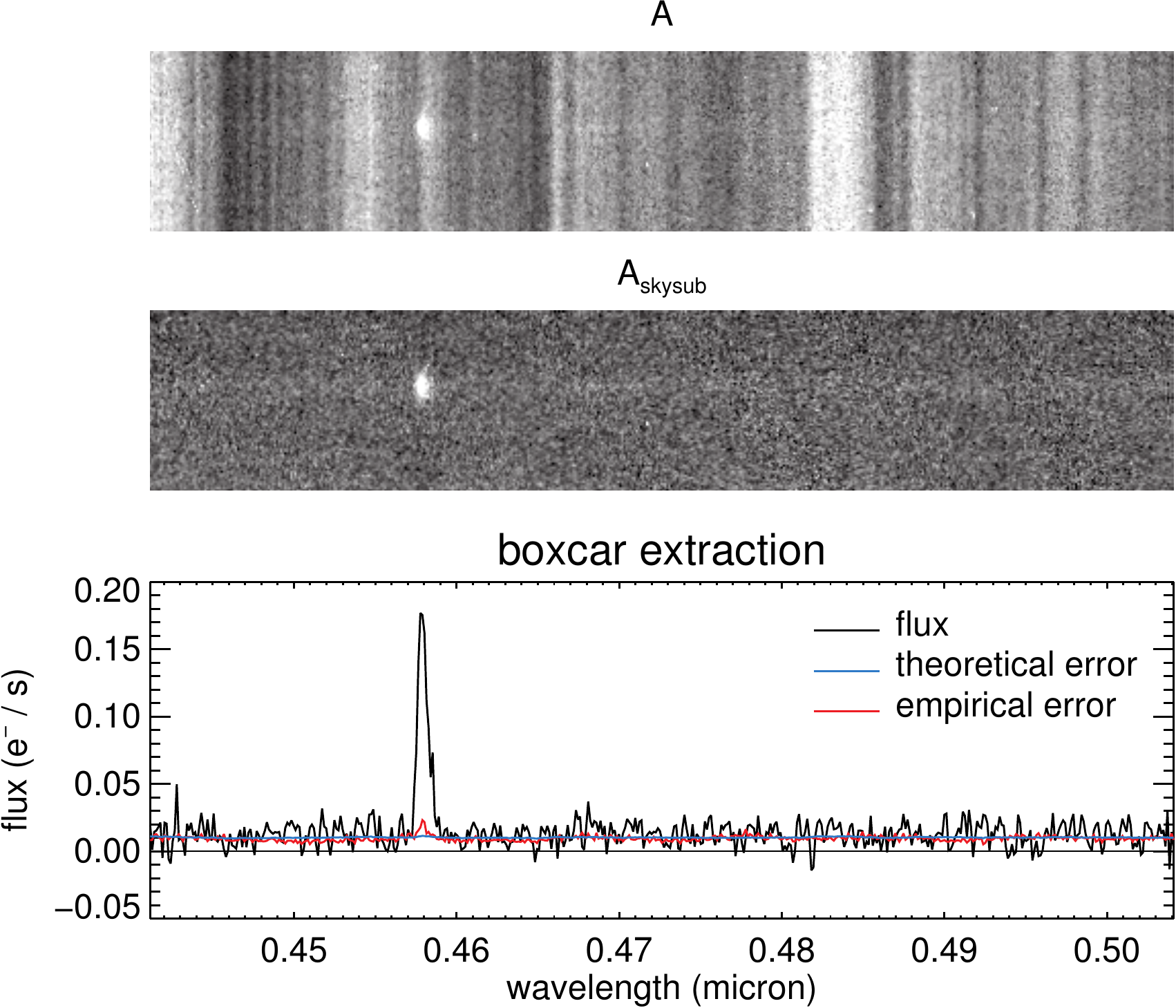}
  \caption{Example of data reduction for observations obtained with the blue arm of LRIS. Top: final stack of five frames (all marked $A$ frames because no nodding was performed), showing faint features from the sky emission and an emission line from a high-redshift galaxy. Only a quarter of  the entire wavelength range of the slit is shown. Center: stack of the sky-subtracted frames. Bottom: boxcar extraction, where the theoretical and the empirical error spectra are also shown.}
   \label{fig:example_LRIS}
\end{figure}

Blue-sensitive CCDs are not heavily affected by cosmic rays, and a brief visual check confirms that the individual frames do not need to be cleaned, despite the long exposure times. One of the most challenging steps for observations covering a very blue spectral range is the tracing of the slit edges, because sky and flat lamps are typically very faint bluewards of $\sim4000\AA$. We found that the most effective way is to use arcs observations, that usually contain a few strong emission lines at blue wavelengths. This works even if the slits are separated by a narrow gap, when the continuum emission in one slit can be difficult to separate from the emission in the the adjacent slits. Because of this choice, we had to manually change the default settings of \flame\ to specify the use of arcs for slit tracing.

Arcs are also used to derive the wavelength calibration, which is then shifted for each frame by fitting the position of [\ion{O}{i}]5577, which is the only bright sky line present in these spectra. The shifts are of the order of a few angstrom, while the difference in the shift measured from different science frames is negligible. This effect is probably due to a mechanical displacement between the arcs frame and the science frames, likely due to flexure. The wavelength coverage falls short of 5577\AA\ for six out of thirty slits; in these cases \flame\ uses a model of the sky spectrum to calculate the spectral shift compared to the arcs wavelength solution. Given the lack of strong, sharp features in the sky spectrum at these wavelengths, this method yields a less accurate result.

Finally, twilight flats are available for these observations, and we adopt them to derive the illumination correction.

Using the same setup as for the LUCI sample observations described above, the data reduction of the 5 frames took 65 minutes for all slits, and only 3 minutes for the complete reduction of one slit.

Given the lack of sky emission lines, it is not possible to verify the goodness of the wavelength calibration on the reduced data. However, an indirect test of the validity of the rectification is given by the high quality of sky subtraction, as shown in Figure \ref{fig:example_LRIS}. As the sky emission at these wavelengths is very faint, it is possible to obtain an excellent sky subtraction that leaves almost no residuals. Similarly to what shown above for LUCI, in this case too the empirical and theoretical estimates of the error spectrum are in remarkable agreement.

% -------------------------------------------------
%  					SUMMARY AND CONCLUSIONS
% -------------------------------------------------

\section{Summary}
\label{sec:summary}

We have presented \flame, a data reduction pipeline for spectroscopic observations obtained with near-infrared or optical instruments. The code is written in IDL and is publicly available. The flexible design of the pipeline ensures compatibility with virtually any singly-dispersed multi-slit spectrograph. Support for new instruments can be easily implemented by developing a simple initialization module that converts the raw data into a standard format and sets appropriate values for the options that control the data reduction steps. Initialization modules for the near-infrared spectrographs LBT/LUCI and Keck/MOSFIRE, and the optical instrument Keck/LRIS are provided, and can be used as a guide for other instruments.

The flexible design is not only ideal for adapting the pipeline to new instruments, but is also well suited for non-standard scientific applications. For example, by choosing to reduce only one of the slits, removing all the calibrations steps, and choosing the fastest but less accurate options, \flame\ can be used very effectively for quick, real-time data reduction while observing.
Another useful application is as a development tool when building new instruments: we plan to use \flame\ to analyze simulated spectroscopic data \citep{leschinski16} for MICADO, one of the first light instruments for the E-ELT \citep{davies16}, and to support the development of the official pipeline.

At the core of the pipeline is the coordinate transformation between the observed and the rectified frame. This is composed of a spatial calibration, which is derived from the automatically traced slit edges, and a wavelength calibration, which can be obtained from either arc or sky emission lines. In addition to isolated emission lines of known wavelength, \flame\ allows users to select spectral features of unknown wavelength, in order to improve the rectification in particularly difficult situations. The algorithms used to derive the wavelength calibration have been developed with the specific goal of being robust, and to work with data at different wavelengths and with varying spectral resolutions. Once the coordinate transformations are known, it is possible to apply illumination correction and sky subtraction without having to resample the data. In fact, the data are resampled only once throughout the data reduction, which minimizes the amount of correlated noise.

Nodding and $A-B$ subtraction, which are typically used in the near-infrared to suppress strong sky lines, are fully supported by \flame. Furthermore, the pipeline takes advantage of the possibility offered by multi-slit instruments of observing a reference star simultaneously with the scientific targets. This allows the observing conditions to be closely tracked, and the nodding pattern to be automatically recognized. In \flame\ the spatial calibration is used not only to correct for geometric distortion in the spectra, but also to account for the spatial misalignment between frames that can be due to either dithering or unwanted drifting. This means that the rectified frames are automatically aligned and can be stacked without further resampling.

We showed, using test data in the near-infrared and in the optical, that \flame\ is able to effectively subtract the sky emission from the science observations. Sky subtraction is effective even for faint targets at near-infrared wavelengths, for which the error spectrum approaches the limit of Poisson noise despite the strong and variable sky lines.

% -------------------------------------------------
%  					ACKNOWLEDGEMENTS
% -------------------------------------------------

\section*{Acknowledgements}

We are grateful to Tom Fletcher for having thoroughly tested the pipeline. We also acknowledge John Stott, Michele Perna, Marco Mignoli, and Dave Thompson. The design of \flame\ was inspired by data reduction code written by many people, including Eva Wuyts, Nick Konidaris and Chuck Steidel.

The LBT is an international collaboration among institutions in the United States, Italy and Germany. LBT Corporation partners are: The University of Arizona on behalf of the Arizona university system; Istituto Nazionale di Astrofisica, Italy; LBT Beteiligungsgesellschaft, Germany, representing the Max-Planck Society, the Astrophysical Institute Potsdam, and Heidelberg University; The Ohio State University, and The Research Corporation, on behalf of The University of Notre Dame, University of Minnesota and University of Virginia.

This work has made use of the Keck Observatory Archive (KOA), which is operated by the W. M. Keck Observatory and the NASA Exoplanet Science Institute (NExScI), under contract with the National Aeronautics and Space Administration.

%%%%%%%%%%%%%%%%%%%%%%%%%%%%%%%%%%%%%%%%%%%%%%%%%%



\bibliographystyle{mnras}
\bibliography{flame}

\begin{thebibliography}{}
\makeatletter
\relax
\def\mn@urlcharsother{\let\do\@makeother \do\$\do\&\do\#\do\^\do\_\do\%\do\~}
\def\mn@doi{\begingroup\mn@urlcharsother \@ifnextchar [ {\mn@doi@}
  {\mn@doi@[]}}
\def\mn@doi@[#1]#2{\def\@tempa{#1}\ifx\@tempa\@empty \href
  {http://dx.doi.org/#2} {doi:#2}\else \href {http://dx.doi.org/#2} {#1}\fi
  \endgroup}
\def\mn@eprint#1#2{\mn@eprint@#1:#2::\@nil}
\def\mn@eprint@arXiv#1{\href {http://arxiv.org/abs/#1} {{\tt arXiv:#1}}}
\def\mn@eprint@dblp#1{\href {http://dblp.uni-trier.de/rec/bibtex/#1.xml}
  {dblp:#1}}
\def\mn@eprint@#1:#2:#3:#4\@nil{\def\@tempa {#1}\def\@tempb {#2}\def\@tempc
  {#3}\ifx \@tempc \@empty \let \@tempc \@tempb \let \@tempb \@tempa \fi \ifx
  \@tempb \@empty \def\@tempb {arXiv}\fi \@ifundefined
  {mn@eprint@\@tempb}{\@tempb:\@tempc}{\expandafter \expandafter \csname
  mn@eprint@\@tempb\endcsname \expandafter{\@tempc}}}

\bibitem[\protect\citeauthoryear{{Chilingarian}, {Beletsky}, {Moran}, {Brown},
  {McLeod}  \& {Fabricant}}{{Chilingarian} et~al.}{2015}]{chilingarian15}
{Chilingarian} I.,  {Beletsky} Y.,  {Moran} S.,  {Brown} W.,  {McLeod} B.,
  {Fabricant} D.,  2015, \mn@doi [\pasp] {10.1086/680598}, \href
  {http://adsabs.harvard.edu/abs/2015PASP..127..406C} {127, 406}

\bibitem[\protect\citeauthoryear{{Davies}}{{Davies}}{2007}]{davies07}
{Davies} R.~I.,  2007, \mn@doi [\mnras] {10.1111/j.1365-2966.2006.11383.x},
  \href {http://adsabs.harvard.edu/abs/2007MNRAS.375.1099D} {375, 1099}

\bibitem[\protect\citeauthoryear{{Davies} et~al.,}{{Davies}
  et~al.}{2013}]{davies13}
{Davies} R.~I.,  et~al., 2013, \mn@doi [\aap] {10.1051/0004-6361/201322282},
  \href {http://adsabs.harvard.edu/abs/2013A%26A...558A..56D} {558, A56}

\bibitem[\protect\citeauthoryear{{Davies} et~al.,}{{Davies}
  et~al.}{2016}]{davies16}
{Davies} R.,  et~al., 2016, in Ground-based and Airborne Instrumentation for
  Astronomy VI. p. 99081Z (\mn@eprint {arXiv} {1607.01954}),
  \mn@doi{10.1117/12.2233047}

\bibitem[\protect\citeauthoryear{{Fruchter} \& {Hook}}{{Fruchter} \&
  {Hook}}{2002}]{fruchter02}
{Fruchter} A.~S.,  {Hook} R.~N.,  2002, \mn@doi [\pasp] {10.1086/338393}, \href
  {http://adsabs.harvard.edu/abs/2002PASP..114..144F} {114, 144}

\bibitem[\protect\citeauthoryear{{Hanuschik}}{{Hanuschik}}{2003}]{hanuschik03}
{Hanuschik} R.~W.,  2003, \mn@doi [\aap] {10.1051/0004-6361:20030885}, \href
  {http://adsabs.harvard.edu/abs/2003A%26A...407.1157H} {407, 1157}

\bibitem[\protect\citeauthoryear{{Horne}}{{Horne}}{1986}]{horne86}
{Horne} K.,  1986, \mn@doi [\pasp] {10.1086/131801}, \href
  {http://adsabs.harvard.edu/abs/1986PASP...98..609H} {98, 609}

\bibitem[\protect\citeauthoryear{{Kelson}}{{Kelson}}{2003}]{kelson03}
{Kelson} D.~D.,  2003, \mn@doi [\pasp] {10.1086/375502}, \href
  {http://adsabs.harvard.edu/abs/2003PASP..115..688K} {115, 688}

\bibitem[\protect\citeauthoryear{{Kriek} et~al.,}{{Kriek}
  et~al.}{2015}]{kriek15}
{Kriek} M.,  et~al., 2015, \mn@doi [\apjs] {10.1088/0067-0049/218/2/15}, \href
  {http://adsabs.harvard.edu/abs/2015ApJS..218...15K} {218, 15}

\bibitem[\protect\citeauthoryear{{Leschinski} et~al.,}{{Leschinski}
  et~al.}{2016}]{leschinski16}
{Leschinski} K.,  et~al., 2016, in Modeling, Systems Engineering, and Project
  Management for Astronomy VI. p. 991124 (\mn@eprint {arXiv} {1609.01480}),
  \mn@doi{10.1117/12.2232483}

\bibitem[\protect\citeauthoryear{{Lord}}{{Lord}}{1992}]{lord92}
{Lord} S.~D.,  1992, NASA Technical Memorandum, 103957

\bibitem[\protect\citeauthoryear{{Markwardt}}{{Markwardt}}{2009}]{markwardt09}
{Markwardt} C.~B.,  2009, in {Bohlender} D.~A.,  {Durand} D.,   {Dowler} P.,
  eds,  Astronomical Society of the Pacific Conference Series Vol. 411,
  Astronomical Data Analysis Software and Systems XVIII. p.~251 (\mn@eprint
  {arXiv} {0902.2850})

\bibitem[\protect\citeauthoryear{{Masters} \& {Capak}}{{Masters} \&
  {Capak}}{2011}]{masters11}
{Masters} D.,  {Capak} P.,  2011, \mn@doi [\pasp] {10.1086/660023}, \href
  {http://adsabs.harvard.edu/abs/2011PASP..123..638M} {123, 638}

\bibitem[\protect\citeauthoryear{{McCarthy} et~al.,}{{McCarthy}
  et~al.}{1998}]{mccarthy98}
{McCarthy} J.~K.,  et~al., 1998, in {D'Odorico} S.,  ed.,  \procspie Vol. 3355,
  Optical Astronomical Instrumentation. pp 81--92, \mn@doi{10.1117/12.316831}

\bibitem[\protect\citeauthoryear{{Oke} et~al.,}{{Oke} et~al.}{1995}]{oke95}
{Oke} J.~B.,  et~al., 1995, \mn@doi [\pasp] {10.1086/133562}, \href
  {http://adsabs.harvard.edu/abs/1995PASP..107..375O} {107, 375}

\bibitem[\protect\citeauthoryear{{Osterbrock}, {Fulbright}, {Martel}, {Keane},
  {Trager}  \& {Basri}}{{Osterbrock} et~al.}{1996}]{osterbrock96}
{Osterbrock} D.~E.,  {Fulbright} J.~P.,  {Martel} A.~R.,  {Keane} M.~J.,
  {Trager} S.~C.,   {Basri} G.,  1996, \mn@doi [\pasp] {10.1086/133722}, \href
  {http://adsabs.harvard.edu/abs/1996PASP..108..277O} {108, 277}

\bibitem[\protect\citeauthoryear{{Osterbrock}, {Fulbright}  \&
  {Bida}}{{Osterbrock} et~al.}{1997}]{osterbrock97}
{Osterbrock} D.~E.,  {Fulbright} J.~P.,   {Bida} T.~A.,  1997, \mn@doi [\pasp]
  {10.1086/133920}, \href {http://adsabs.harvard.edu/abs/1997PASP..109..614O}
  {109, 614}

\bibitem[\protect\citeauthoryear{{Rockosi} et~al.,}{{Rockosi}
  et~al.}{2010}]{rockosi10}
{Rockosi} C.,  et~al., 2010, in Ground-based and Airborne Instrumentation for
  Astronomy III. p. 77350R, \mn@doi{10.1117/12.856818}

\bibitem[\protect\citeauthoryear{{Rousselot}, {Lidman}, {Cuby}, {Moreels}  \&
  {Monnet}}{{Rousselot} et~al.}{2000}]{rousselot00}
{Rousselot} P.,  {Lidman} C.,  {Cuby} J.-G.,  {Moreels} G.,   {Monnet} G.,
  2000, \aap, \href {http://adsabs.harvard.edu/abs/2000A%26A...354.1134R} {354,
  1134}

\bibitem[\protect\citeauthoryear{{Seifert} et~al.,}{{Seifert}
  et~al.}{2003}]{seifert03}
{Seifert} W.,  et~al., 2003, in {Iye} M.,  {Moorwood} A.~F.~M.,  eds,
  \procspie Vol. 4841, Instrument Design and Performance for Optical/Infrared
  Ground-based Telescopes. pp 962--973, \mn@doi{10.1117/12.459494}

\bibitem[\protect\citeauthoryear{{Soto}, {Lilly}, {Bacon}, {Richard}  \&
  {Conseil}}{{Soto} et~al.}{2016}]{soto16}
{Soto} K.~T.,  {Lilly} S.~J.,  {Bacon} R.,  {Richard} J.,   {Conseil} S.,
  2016, \mn@doi [\mnras] {10.1093/mnras/stw474}, \href
  {http://adsabs.harvard.edu/abs/2016MNRAS.458.3210S} {458, 3210}

\bibitem[\protect\citeauthoryear{{Steidel}, {Shapley}, {Pettini}, {Adelberger},
  {Erb}, {Reddy}  \& {Hunt}}{{Steidel} et~al.}{2004}]{steidel04}
{Steidel} C.~C.,  {Shapley} A.~E.,  {Pettini} M.,  {Adelberger} K.~L.,  {Erb}
  D.~K.,  {Reddy} N.~A.,   {Hunt} M.~P.,  2004, \mn@doi [\apj]
  {10.1086/381960}, \href {http://adsabs.harvard.edu/abs/2004ApJ...604..534S}
  {604, 534}

\bibitem[\protect\citeauthoryear{{Steidel} et~al.,}{{Steidel}
  et~al.}{2014}]{steidel14}
{Steidel} C.~C.,  et~al., 2014, \mn@doi [\apj] {10.1088/0004-637X/795/2/165},
  \href {http://adsabs.harvard.edu/abs/2014ApJ...795..165S} {795, 165}

\bibitem[\protect\citeauthoryear{{Vacca}, {Cushing}  \& {Rayner}}{{Vacca}
  et~al.}{2003}]{vacca03}
{Vacca} W.~D.,  {Cushing} M.~C.,   {Rayner} J.~T.,  2003, \mn@doi [\pasp]
  {10.1086/346193}, \href {http://adsabs.harvard.edu/abs/2003PASP..115..389V}
  {115, 389}

\bibitem[\protect\citeauthoryear{{Weilbacher}, {Streicher}, {Urrutia}, {Jarno},
  {P{\'e}contal-Rousset}, {Bacon}  \& {B{\"o}hm}}{{Weilbacher}
  et~al.}{2012}]{weilbacher12}
{Weilbacher} P.~M.,  {Streicher} O.,  {Urrutia} T.,  {Jarno} A.,
  {P{\'e}contal-Rousset} A.,  {Bacon} R.,   {B{\"o}hm} P.,  2012, in Software
  and Cyberinfrastructure for Astronomy II. p. 84510B,
  \mn@doi{10.1117/12.925114}

\bibitem[\protect\citeauthoryear{{van Dokkum}}{{van
  Dokkum}}{2001}]{vandokkum01}
{van Dokkum} P.~G.,  2001, \mn@doi [\pasp] {10.1086/323894}, \href
  {http://adsabs.harvard.edu/abs/2001PASP..113.1420V} {113, 1420}

\makeatother
\end{thebibliography}

%%%%%%%%%%%%%%%%%%%%%%%%%%%%%%%%%%%%%%%%%%%%%%%%%%

%%%%%%%%%%%%%%%%% APPENDICES %%%%%%%%%%%%%%%%%%%%%

\appendix

\section{An Example of Driver File}
\label{sec:driverfile}

The reduction of a data set with \flame\ is entirely controlled via a \emph{driver file}. This is a simple text file containing a series of IDL commands that can be run either automatically or line-by-line in an interactive session. We report here the content of an example driver file:

\begin{verbatim}
  input = flame_create_input()
  input.science_filelist = 'science.txt'
  input.pixelflat_filelist = 'flats.txt'
  input.dark_filelist = 'darks.txt'
  input.AB_subtraction = 1
  input.star_y_A = 676
  input.star_y_B = 662
  input.reduce_only_oneslit = 0

  fuel = flame_initialize_luci(input)
  fuel.settings.skysub_reject_fraction = 0.05
  fuel.settings.skysub_reject_loops = 8

  flame_diagnostics, fuel
  flame_quickstack, fuel
  flame_calibrations, fuel
  flame_slitid, fuel
  flame_cutouts, fuel
  flame_spatialcal, fuel
  flame_roughwavecal, fuel
  flame_findlines, fuel
  flame_wavecal, fuel
  flame_illumcorr, fuel
  flame_skysub, fuel
  flame_rectify, fuel
  flame_combine, fuel
  flame_checkdata, fuel
  flame_extract, fuel
\end{verbatim}

The user needs to edit the first part of the file where the \texttt{input} structure is defined. The most important inputs are the names of the text files containing the list of science and calibration frames; a flag indicating whether $A-B$ nodding was performed; and the approximate vertical pixel coordinate of the trace of the reference star. The \texttt{input} structure contains more optional fields that can be specified when needed, for example when reducing longslit data, using arcs or slit flats, adopting manual nodding values, and so on.

Once the inputs are set, the \texttt{fuel} structure, containing all the information needed for reducing a data set, is initialized. The initialization routine is unique to each instrument, and sets the various parameters that control each of the data reduction steps. In order to add support for a new instrument, it is sufficient to write a new initialization module.

After the initialization, the \flame\ settings can be further changed by the user for specific applications. In the driver file shown above, for example, the parameters controlling the outlier rejection when fitting a model of the sky spectrum have been manually changed to a rejection of 5\% of the pixels for eight consecutive times. This ensures a slower, but more thorough, filtering of potential emission from astronomical sources compared to the default settings, and is appropriate when the target occupies a large fraction of the slit along the spatial direction.

Finally, the last part of the driver file consists of a series of commands that perform the various steps of the data reduction, and should normally not be modified by the user. Each command performs an individual step, and takes the \texttt{fuel} structure as its only argument. If a set of data requires a specific treatment which is not implemented in \flame, the user can easily edit or replace one or more of these routines, leaving the structure of the data reduction pipeline unchanged.

%%%%%%%%%%%%%%%%%%%%%%%%%%%%%%%%%%%%%%%%%%%%%%%%%%

% Don't change these lines
%\bsp	% typesetting comment
\label{lastpage}
\end{document}